%% file: main.tex
\newif\iffinal
  \finaltrue

\documentclass{eptcs}

\iffinal\else{
  \IfFileExists{crop_pages}{
  \paperwidth 13cm
  \paperheight 20cm
  \oddsidemargin  -0.85in
  \evensidemargin -0.85in
  \topmargin      -1.25in
  }{}
\fi

\input{macros}  % some useful commands

\title{\CTLstar Synthesis \,via\, \LTL Synthesis}

\author{Roderick Bloem$^1$, Sven Schewe$^2$, Ayrat Khalimov$^1$
        \institute{%
        $^1$ Graz University of Technology, Austria\\
        $^2$ University of Liverpool, UK}
        \thanks{The authors-order was decided by tossing the coin.}}

\def\titlerunning{\CTLstar Synthesis via LTL Synthesis}
\def\authorrunning{R. Bloem, S. Schewe, A. Khalimov}

\begin{document}

\maketitle

\begin{abstract}
We reduce synthesis for \CTLstar properties to synthesis for \LTL.
In the context of model checking this is impossible --- \CTLstar is more expressive than LTL.
Yet, in synthesis we have knowledge of the system structure \emph{and} we can add new outputs.
These outputs can be used to encode witnesses of the satisfaction of \CTLstar subformulas directly into the system.
This way, we construct an LTL formula, over old and new outputs and original inputs,
which is realisable if, and only if, the original \CTLstar formula is realisable.
The \CTLstar-via-LTL synthesis approach preserves the problem complexity,
although it might increase the minimal system size.
We implemented the reduction,
and evaluated the \CTLstar-via-LTL synthesiser on several examples.
\end{abstract}

\iffinal\else
\textcolor{blue}{\small{
TODOs:
\li
\- argh, path formula can be \CTLstar, while LTL formula is LTL!
   Check if the main section correctly uses them (I changed at some point..)
\- sync defs with the main
\- related work
\- unrealisability section: elaborate on the witness
\il
}
}
\fi

\section{Introduction} \label{sec:intro}

In reactive synthesis we automatically construct a system from a given specification in some temporal logic.
The problem was introduced by Church for Monadic Second Order Logic~\cite{Church63}.
Later Pnueli introduced Linear Temporal Logic (LTL)~\cite{pnueli1977temporal}
and together with Rosner proved 2EXPTIME-completeness
of the reactive synthesis problem for LTL~\cite{DBLP:conf/popl/PnueliR89}.
In parallel, Emerson and Clarke introduced Computation Tree Logic (CTL)~\cite{ctl-origin},
and later Emerson and Halpern introduce Computation Tree Star Logic (\CTLstar)~\cite{ctlstar-origin}
that subsumes both CTL and LTL.
Kupferman and Vardi showed~\cite{informatio} that the synthesis problem for \CTLstar is 2EXPTIME-complete.
%The original approach to synthesis from LTL goes via Safra construction~\cite{DBLP:conf/focs/Safra88,DBLP:conf/lics/Piterman06}
%to determinize B\"uchi automata \cite{DBLP:conf/popl/PnueliR89},
%which proved to be hard to implement efficiently.

Intuitively, LTL allows one to reason about infinite computations.
The logic has \emph{temporal} operators, e.g., $\G$ (always) and $\F$ (eventually),
and allows one to state properties like ``every request is eventually granted''
($\G(r \impl \F g)$).
A system satisfies a given LTL property if \emph{all} its computations satisfy it.

In contrast, CTL and \CTLstar reason about computation trees,
usually derived by unfolding the system.
The logics have---in addition to temporal operators---\emph{path quantifiers}:
$\A$ (on all paths) and $\E$ (there exists a path).
CTL forbids arbitrary nesting of path quantifiers and temporal operators:
they must interleave.
E.g.,
$\AG g$ (``on all paths we always grant'') is a CTL formula,
but $\AGF g$ (``on all paths we infinitely often grant'') is not a CTL formula.
\CTLstar lifts this limitation.

The expressive powers of CTL and LTL are incomparable:
there are systems indistinguishable by CTL but distiniguishable by LTL, and vice versa.
One important property inexpressible in LTL is the resettability property:
``there is always a way to reach the `reset' state'' ($\AGEF reset$).

% In synthesis,
% \CTLstar also allows the designer to write ``cooperative'' properties,
% that say that a specific behaviour is possible when the environment cooperates,
% yet it leaves open how the environment should cooperate.
% The synthesiser then needs to find a system and the behaviour of the environment
% that together satisfy the property.
% Existentially quantified properties can also be used to ensure that
% the system does not synthesise the specification vacuously, by falsifying the assumptions.

There was a time when CTL and LTL competed for ``best logic for model checking''~\cite{LTL-vs-CTL}.
Nowadays most model checkers use LTL.
LTL is also prevalent in reactive synthesis.
SYNTCOMP~\cite{syntcomp}---the reactive synthesis competetion with the goal to popularise reactive synthesis---%
has two distinct tracks, and both use LTL as their specification language.

Yet LTL leaves the designer without \emph{structural} properties.
One solution is to develop general \CTLstar synthesisers like the one in~\cite{CTLstarCAV}.
Another solution is to transform the \CTLstar synthesis problem into
the form understandable to LTL synthesisers, i.e., to reduce \CTLstar synthesis to LTL synthesis.
Such a reduction would automatically transfer performance advances in LTL synthesisers
to a \CTLstar synthesiser.
This paper shows one such reduction.

Our reduction of \CTLstar synthesis to LTL synthesis works as follows.

First, recall how the standard \CTLstar model checking works~\cite{PrinciplesMC}.
The verifier introduces a proposition for every state subformula---formulas starting with an $\A$ or an $\E$ path quantifier---of a given \CTLstar formula.
Then the verifier annotates system states with these propositions,
in the bottom up fashion,
starting with propositions that describe subformulas over original propositions (system inputs and outputs) only.
Therefore the system satisfies the \CTLstar formula iff the initial system state is annotated
with the proposition describing the whole \CTLstar formula
(assuming that the \CTLstar formula starts with $\A$ or $\E$).

Now let us look into \CTLstar synthesis.
The synthesiser has the flexibility to choose the system structure, 
as long as it satisfies a given specification.
We introduce new propositions---outputs that later can be hidden from the user---%
for state subformulas of the \CTLstar formula,
just like in the model checking case above.
We also introduce additional propositions for existentially quantified subformulas---%
to encode the witnesses of their satisfaction.
Such propositions describe the directions (inputs) the system should take
to satisfy existentially quantified path formulas.
The requirement that new propositions indeed denote the truth of the subformulas can be stated in LTL.
For example, for a state subformula $\A\varphi$, we introduce proposition $p_{\A\varphi}$,
and require $\G\left[ p_{\A\varphi} \impl \varphi' \right]$,
where $\varphi'$ is $\varphi$ with state subformulas substituted by the propositions.
For an existential subformula $\E\varphi$,
we introduce proposition $p_{\E\varphi}$ and require,
\emph{roughly}, $\G\left[ p_{\E\varphi} \impl ((\G d_{p_{\E\varphi}}) \impl \varphi')\right]$, which states:
if the proposition $p_{\E\varphi}$ holds, then the path along directions encoded by $d_{p_{\E\varphi}}$
satisfies $\varphi'$ (where $\varphi'$ as before).
We wrote ``roughly'', because
there can be several different witnesses for the same existential subformula
starting at different system states:
they may meet in the same system state,
but depart afterwards---then, to able to depart from the meeting state,
each witness should have its own direction $d$.
We show that, for each existential subformula, a number $\approx 2^{|\Phi_\CTLstar|}$ of witnesses is sufficient,
where $\Phi_\CTLstar$ is a given \CTLstar formula.
This makes the LTL formula exponential in the size of the \CTLstar formula,
but the special---conjunctive---nature of the LTL formula ensures
that the synthesis complexity is 2EXPTIME wrt. $|\Phi_\CTLstar|$.

Our reduction is ``if and only if'', and it preserves the synthesis complexity.
However, it may increase the size of the system, and is not very well suited to establish unrealisability.
Of course, to show that the \CTLstar formula is unrealisable,
one could reduce \CTLstar synthesis to LTL synthesis,
then reduce the LTL synthesis problem to solving parity games,
and derive the unrealisability from there%
\footnote{Reducing LTL synthesis to solving parity games \emph{is} practical, as SYNTCOMP'17~\cite{syntcomp} showed:
  such synthesiser {\tt ltlsynt} was among the fastest.}.
But the standard approach for unrealisability checking---by synthesising the dualised LTL specification---does not seem to be practical,
since the automaton for the negated LTL formula explodes in size.

Finally, we have implemented\footnote{Available at \url{https://github.com/5nizza/party-elli}, branch ``cav17''}
the converter from \CTLstar into LTL,
and evaluated \CTLstar-via-LTL synthesis approach,
using two LTL synthesisers and \CTLstar synthesiser~\cite{CTLstarCAV}, on several examples.
The experimental results show that such an approach works very well%
---outperforming the specialised \CTLstar synthesiser~\cite{CTLstarCAV}---%
when the number of \CTLstar-specific formulas is small.

The paper structure is as follows.
Section~\ref{sec:defs} defines
B\"uchi and co-B\"uchi word automata, tree automata,
\CTLstar with inputs,
Moore systems, computation trees, and other useful notions.
Section~\ref{sec:reductions-to-ltl} contains the main contribution: it describes the reduction.
In Section~\ref{sec:ctlstar-unreal} we briefly discuss checking unrealisability of \CTLstar specifications.
Section~\ref{sec:experiments} describes the experimental setup, specifications, solvers used, and synthesis timings.
We conclude in Section~\ref{sec:conclusion}.
\ak{mention related work section}

\section{Definitions} \label{sec:defs}

Notation:
$\bbB = \{\true,\false\}$ is the set of Boolean values,
$\bbN$ is the set of natural numbers (excluding $0$),
$[i,j]$ for integers $i \leq j$ is the set $\{i,...,j\}$,
$[k]$ is $[1,k]$ for $k \in \bbN$.
By default, we use natural numbers.
%Also, for an arbitrary set $I$,
%the calligraphic writing $\cal I$ denotes $2^I$.

In this paper we consider \emph{finite} systems and automata.

\subsection{Moore Systems}

A \emph{(Moore) system} $M$ is a tuple
$(I, O, T, t_0, \tau, out)$
where
$I$ and $O$ are disjoint sets of input and output variables,
$T$ is the set of states, $t_0 \in T$ is the initial state,
$\tau: T \times \I \to T$ is a transition function,
$out: T \to \O$ is the output function that
labels each state with a set of output variables.
Note that systems have no dead ends and have a transition for every input.
We write $t \trans{io} t'$ when $t' = \tau(t,i)$ and $out(t) = o$.

For the rest of the section, fix a system $M=(I, O, T, t_0, \tau, out)$.

A \emph{system path} is a sequence $t_1 t_2 ... \in T^\omega$
such that, for every $i$, there is $e \in \I$ with $\tau(t_i,e) = t_{i+1}$.
An \emph{input-labeled system path} is a sequence $(t_1,e_1) (t_2,e_2) ... \in (T\times \I)^\omega$
where $\tau(t_i,e_i) = t_{i+1}$ for every $i$.
A \emph{system trace starting from $t_1 \in T$} is a sequence $(o_1\cup e_1) (o_2\cup e_2) ... \in (\I \cup \O)^\omega$,
for which there exists an input-labeled system path $(t_1,e_1) (t_2,e_2) ...$ 
and $o_i=out(t_i)$ for every $i$.
Note that, since systems are Moore,
the output $o_i$ cannot ``react'' to input $e_i$.
I.e.,
the outputs are ``delayed'' with respect to inputs.

\subsection{Trees}

A \emph{(infinite) tree} is a tuple $(D, L, V \subseteq D^*, l:V \to L)$,
where
\li
\- $D$ is the set of directions,
\- $L$ is the set of node labels,
\- $V$ is the set of nodes satisfying:
   (i) $\epsilon \in V$ is called the root (the empty sequence),
  (ii) $V$ is closed under prefix operation (i.e., every node is connected to the root),
 (iii) for every $n \in V$ there exists $d \in D$ such that $n\cdot d \in V$
       (i.e., there are no leafs),
\- $l$ is the nodes labeling function.
\il
A tree $(D,L,V,l)$ is \emph{exhaustive} iff $V=D^*$.
%A tree is \emph{non-labeled} iff $|L|=1$ and then we omit $L$ and $l$.

A \emph{tree path} is a sequence $n_1 n_2 ... \in V^\omega$,
such that, for every $i$, there is $d \in D$ and $n_{i+1} = n_i \cdot d$.

%An \emph{$L$-labeled $D$-directed tree} is a tuple $(V,l)$, where
%$V=D^*$ is the (infinite) set of tree nodes
%and $\epsilon \in V$ (the empty sequence) is called the root node,
%$l: V \to L$ is a labeling function.
%An \emph{(infinite) tree path} is a sequence $n_1 n_2 ... \in V^\omega$,
%such that, for every $i$, there is $d \in D$ and $n_{i+1} = n_i \cdot d$.

In contexts where $I$ and $O$ are inputs and outputs,
we call an exhaustive tree $(D=\I, L=\O, V=D^*, l:V \to \O)$
a \emph{computation tree}.
We omit $D$ and $L$ when they are clear from the context.
E.g. we can write $(V=(2^I)^*, l:V\to2^O)$ instead of
$(\I, \O, V=(2^I)^*, l:V \to \O)$.

With every system $M=(I, O, T, t_0, \tau, out)$ we associate
the computation tree $(D, L, V, l)$ such that, for every $n\in V$:
$l(n)=out(\tau(t_0,n))$,
where $\tau(t_0,n)$ is the state, in which the system,
starting in the initial state $t_0$,
ends after reading the input word $n$.
We call such a tree a \emph{system computation tree}.

A computation tree is \emph{regular}
iff it is a system computation tree for some system.

\subsection{\CTLstar with Inputs (release PNF) and LTL}\label{sec:def:ctlstar}
\rb{why is it called \emph{release} PNF?}

For this section, fix two disjoint sets: inputs $I$ and outputs $O$.
Below we define \CTLstar with inputs (in release positive normal form).
The definition differentiates inputs and outputs (see Remark~\ref{rem:ctlstar-subtle}).

\parbf{Syntax of \CTLstar with inputs}
\emph{State formulas} have the grammar:
$$
\Phi = \true \| \false \|
       o \| \neg o \| \Phi \land \Phi \| \Phi \lor \Phi \|
       \A \varphi \| \E \varphi
$$
where $o \in O$ and $\varphi$ is a path formula. \emph{Path formulas} are defined by the grammar:
$$
\varphi = \Phi \|
      i \| \neg i \|
      \varphi \land \varphi \| \varphi \lor \varphi \|
      \X \varphi \|
      \varphi \U \varphi \|
      \varphi \R \varphi,
$$
where $i \in I$.
The temporal operators $\G$ and $\F$ are defined as usual.

The above grammar describes the \CTLstar formulas in positive normal form.
The general \CTLstar formula
(in which negations can appear anywhere)
can be converted into the formula of this form with no size blowup,
using equivalence $\neg (a \U b) \equiv \neg a \R \neg b$.

\parbf{Semantics of \CTLstar with inputs}
We define the semantics of \CTLstar with respect to a computation tree
$(V,l)$.
The definition is very similar to the standard one~\cite{PrinciplesMC},
except for a few cases involving inputs
(marked with ``+'').

Let $n \in V$ and $o \in O$.
Then:
\li
\- $n \not\models \Phi$ iff $n \models \Phi$ does not hold
\- $n \models \true$ and $n \not\models \false$
\- $n \models o$ iff $o \in l(n)$, $n \models \neg o$ iff $o \not\in l(n)$
\- $n \models \Phi_1 \land \Phi_2$ iff $n \models \Phi_1$ and $n \models \Phi_2$.
   Similarly for $\Phi_1\lor\Phi_2$.
\-[+] $n \models \A \varphi$ iff for all tree paths $\pi$ starting from $n$:
   $\pi \models \varphi$.
   For $\E\varphi$, replace ``for all'' with ``there exists''.
\il

Let $\pi = n_1 n_2 ... \in V^\omega$ be a tree path,
$i \in I$, and $n_2 = n_1 \cdot e$ where $e \in 2^I$.
For $k \in \bbN$, define $\pi_{[k:]} = n_k n_{k+1} ...$,
i.e., the suffix of $\pi$ starting in $n_k$.
Then:
\li
\- $\pi \models \Phi$ iff $n_1 \models \Phi$
\-[+] $\pi \models i$ iff $i \in e$,
      $\pi \models \neg i$ iff $i \not\in e$.
      Note how inputs are shifted wrt. outputs.
\- $\pi \models \varphi_1 \land \varphi_2$ iff $\pi \models \varphi_1$ and $\pi \models \varphi_2$.
   Similarly for $\varphi_1 \lor \varphi_2$.
\- $\pi \models \X \varphi$ iff $\pi_{[2:]} \models \varphi$
\- $\pi \models \varphi_1 \U \varphi_2$ iff
   $\exists l\in\bbN: (\pi_{[l:]} \models \varphi_2 \land \forall m \in [1,l-1]: \pi_{[m:]} \models \varphi_1)$
\- $\pi \models \varphi_1 \R \varphi_2$ iff
   $(\forall l\in\bbN: \pi_{[l:]} \models \varphi_2) \lor 
    (\exists l\in\bbN: \pi_{[l:]} \models \varphi_1 \land \forall m\in [1,l]: \pi_{[m:]} \models \varphi_2)$
\il

A \emph{computation tree $(V,l)$ satisfies a \CTLstar state formula $\Phi$},
written $(V,l) \models \Phi$,
iff the root node satisfies it.
A \emph{system $M$ satisfies a \CTLstar state formula $\Phi$},
written $M \models \Phi$,
iff its computation tree satisfies it.

\begin{remark}[Subtleties]\label{rem:ctlstar-subtle}
Note that $(V,l) \models i\land o$ is not defined,
since $i \land o$ is not a state formula.
Let $r \in I$ and $g \in O$.
By the semantics, $\E r \equiv \true$ and $\E \neg r \equiv \true$,
while $\E g \equiv g$ and $\E \neg g \equiv \neg g$.
This are the consequences of how we group inputs with outputs.
\end{remark}

\parbf{LTL}
The syntax of LTL formula (in general form) is:
$$
\phi = \true \| \false \| p \| \neg p \| \phi \land \phi \| \neg \phi \| \phi \U \phi \| \X \phi,
$$
where $p \in I \cup O$.
Temporal operators $\G$ and $\F$ are defined as usual.
The semantics is standard (see e.g. \cite{PrinciplesMC}),
and can be derived from that of \CTLstar
assuming that $\pi \models \neg\phi$ iff $\pi\not\models\phi$.
A computation tree $(V,l)$ satisfies an LTL formula $\phi$,
written $(V,l) \models \phi$,
iff all tree paths starting in the root satisfy it.
A system satisfies an LTL formula iff its computation tree satisfies it.

\subsection{Word Automata}

A \emph{word automaton} $A$ is a tuple $(\Sigma, Q, q_0, \delta, acc)$
where $\Sigma$ is an alphabet, $Q$ is a set of states, $q_0 \in Q$ is the initial state,
$\delta: Q \times \Sigma \to 2^Q\smallsetminus \{\emptyset\}$ is a transition relation,
$acc:Q^\omega \to \bbB$ is a path acceptance condition.
Note that word automata have no dead ends and have a transition for every letter of the alphabet.
A word automaton is \emph{deterministic} when $|\delta(q,\sigma)|=1$ for every $(q,\sigma) \in Q\times \Sigma$.
% ak: actually we do not assume automata to be complete

For the rest of this section,
fix word automaton $A=(\Sigma, Q, q_0, \delta, acc)$
with $\Sigma=2^{I \cup O}$.

\emph{A path in automaton $A$}
is a sequence $q_1 q_2 ... \in Q^\omega$ such that
there exists $a_i \in \Sigma$ for every $i$
such that $(q_i,a_i,q_{i+1}) \in \delta(q_i)$.
\emph{A word $a_1 a_2 ...  \in \Sigma^\omega$ generates a path}
$\pi = q_1\dots $ iff for every $i$: $(q_i,a_i,q_{i+1}) \in \delta$.
A \emph{path $\pi$ is accepted} iff $acc(\pi)$ holds.

We define two acceptance conditions.
Let $\pi \in Q^\omega$,
$\Inf(\pi)$ be the elements of $Q$ appearing in $\pi$ infinitely often,
and $F \subseteq Q$.
Then:
\li
\- \emph{B\"uchi acceptance}:
   $acc(\pi)$ holds iff $\Inf(\pi) \cap F \neq \emptyset$.
\- \emph{co-B\"uchi acceptance}:
   $acc(\pi)$ holds iff $\Inf(\pi) \cap F = \emptyset$.
\il

We distinguish two types of word automata: universal and non-deterministic ones.
A \emph{nondeterministic word automaton $A$ accepts a word} from $\Sigma^\omega$
iff there exists an accepted path generated by the word that starts in an initial state.
Universal word automata require \emph{all} such paths to be accepted.
%We write $L(A)$ for the set of all infinite words accepted by $A$.\ak{not used}
%Equvalently,
%in both cases:
%an automaton accepts an infinite sequence iff there is an accepting run.

%We distinguish between two \emph{path quantifiers}, \E and \A:
%$M \models \E(A)$ iff there is a system trace
%$(o_1\cup e_1)(o_2 \cup e_2)...$ that
%starts in the initial system state and is accepted by the automaton;
%$M \models \A(A)$ iff every such system trace is accepted by the automaton.

%The \emph{product $M\times A$}\ak{used?}
%is the automaton $(Q\times T, Q_0 \times \{t_0\}, \Delta, acc')$ such that
%for all $(q,t) \in Q\times T$:
%$\Delta(q,t)=\{(\delta(q,i\cup out(t)),\tau(q,i)) \| i \in \I\} $.
%Define $acc'$ to return true for a given $\pi \in (Q \times T)^\omega$
%iff
%$acc$ returns true for the corresponding projection of $\pi$ into $Q$.
%Note that $M \times A$ has the 1-letter alphabet (not shown in the tuple).
%\ak{note observations about relation of product traces and $M\models \A(UXW)$ and $M \models \E(NXW)$}
%

\parbf{Abbreviations}
NBW means nondeterministic B\"uchi automaton, and
UCW means universal co-B\"uchi automaton.

\subsection{Synthesis Problem}
The \emph{\CTLstar synthesis problem} is:

\smallskip
\noindent
\emph{%
Given: the set of inputs $I$, the set of outputs $O$, \CTLstar formula $\Phi$\\
Return: a computation tree satisfying $\Phi$, otherwise ``unrealisable''
}

\smallskip
\noindent
The inputs to the problem are called a \emph{specification}.
A specification is realisable if the answer is a tree,
and then the tree is called a \emph{model} of the specification.
Similarly we can define the LTL synthesis problem.

It is known~\cite{informatio,DBLP:conf/popl/PnueliR89} that
the \CTLstar and LTL synthesis problems are 2EXPTIME-complete,
and any realisable specification has a regular computation tree model.

\subsection{Tree Automata}
This paper can be understood without
complete understanding of alternating tree automata,
but since they are mentioned in several places,
we define them here.
Namely, below we define alternating hesitant tree automata~\cite{ATA},
which describe \CTLstar formulas,
similarly to how NBWs describe LTL formulas.
The difference is due to the mix of \E and \A path quantifiers---%
hesitant tree automata have an acceptance condition that mixes
B\"uchi and co-B\"uchi acceptance conditions and certain structural properties.

We start with a general case of alternating tree automata
and then define alternating hesitant tree automata.

For a finite set $S$,
let ${\cal B}^+(S)$ denote the set of all positive Boolean formulas over elements of $S$.
%Note that $\false \not\in {\cal B}^+(S)$.

\subsection*{Alternating Tree Automata}

An \emph{alternating tree automaton} is a tuple
$(\Sigma, D, Q, q_0, \delta, acc)$,
where $\Sigma$ is the set of node propositions,
$D$ is the set of directions, $q_0 \subseteq Q$ is the initial state,
$\delta: Q \times \Sigma \to \mathcal{B}^+(D \times Q)$
is the transition relation,
and $acc$ is an acceptance condition $acc: Q^\omega \to \bbB$.
Note that $\delta(q,\sigma) \neq \false$ for every $(q,\sigma) \in Q\times\Sigma$,
i.e., there is always a transition.
Tree automata consume exhaustive trees like $(D, L=\Sigma, V=D^*, l:V \to \Sigma)$
and produce run-trees.

Fix two disjoint sets, inputs $I$ and outputs $O$.

\emph{Run-tree} of an alternating tree automaton
$(\Sigma=2^O, D=2^I, Q, q_0, \delta, acc)$
on a computation tree $(V=(2^I)^*, l:V \to \O)$
is a tree
with directions $2^I \times Q$,
labels $V \times Q$,
nodes $V' \subseteq (2^I \times Q)^*$,
labeling function $l'$
such that
\li
\- $l'(\epsilon) = (\epsilon, q_0)$,
\- if $v \in V'$ with $l'(v) = (n,q)$, then:\\
   there exists $\{(d_1,q_1),...,(d_k,q_k)\}$ that satisfies $\delta(q, l(n))$
   and $n \cdot (d_i, q_i) \in V'$ for every $i \in [1,k]$.
\il
Intuitively,
we run the alternating tree automaton on the computation tree:
\li
\-[(1)]
   We mark the root node of the computation tree with the automaton initial state $q_0$.
   We say that initially, in the node $\epsilon$,
   there is only one copy of the automaton and it has state $q_0$.
\-[(2)]
   We read the label $l(n)$ of the current node $n$ of the computation tree
   and consult the transition function $\delta(q,l(n))$.
   The latter gives a set of conjuncts of atoms of the form $(d',q') \in D\times Q$.
   We nondeterministically choose one such conjunction $\{(d_1,q_1), ..., (d_k,q_k)\}$
   and send a copy of the alternating automaton into each direction $e_i$ in the state $q_i$.
   Note that we can send up to $|Q|$ copies of the automaton into one direction
   (but into different automaton states).
   That is why a run-tree defined above has directions $2^I\times Q$
   rather than $2^I$.
\-[(3)]
   We repeat step (2) for every copy of the automaton.
   As a result we get a run-tree:
   the tree labeled with nodes of the computation tree and
   states of the automaton.
\il

A \emph{run-tree is accepting}
iff every run-tree path starting from the root is accepting.
A run-tree path $v_1 v_2 ...$ is accepting
iff $acc(q_1q_2...)$ holds ($acc$ is defined later),
where $q_i$ for every $i\in \bbN$ is the automaton state part of $l'(v_i)$.

An \emph{alternating tree automaton $A=(\Sigma=2^O, D=2^I, Q, q_0, \delta, acc)$
accepts a computation tree $(V=(2^I)^*, l:V \to \O)$},
written $(V,l) \models A$,
iff
the automaton has an accepting run-tree on that computation tree.
An alternating tree automaton is \emph{non-empty} iff there exists a computation tree accepted by it.

Similarly,
\emph{a Moore system $M=(I, O, T, t_0, \tau, out)$
is accepted
by the alternating tree automaton $A=(\Sigma=2^O, D=2^I, Q, q_0, \delta, acc)$},
written $M \models A$,
iff $(V,l) \models A$,
where $(V=(2^I)^*,l:V\to 2^O)$ is the system computation tree.

Different variations of acceptance conditions are defined the same way as for word automata.

We can define nondeterministic and universal tree automata
in a way similar to word automata.

\subsection*{Alternating Hesitant Tree Automata (AHT)} \label{page:defs:aht}

An \emph{alternating hesitant tree automaton (AHT)} is an alternating tree automaton
$(\Sigma, D, Q, q_0, \delta, acc)$
with the following acceptance condition and structural restrictions.
The restrictions reflect the fact that AHTs are tailored for \CTLstar formulas.
\li 
\- $Q$ can be partitioned into $Q^N_1,\dots ,Q^N_{k_N}$, $Q^U_1,\dots
,Q^U_{k_U}$, where superscript $N$ means
nondeterministic and $U$ means universal.
Let $Q^N = \bigcup Q^N_i$ and $Q^U = \bigcup Q^U_i$.
(Intuitively,
 nondeterministic state sets describe \E-quantified subformulas of the \CTLstar formula,
 while universal --- \A-quantified subformulas.)

\- There is a partial order on $\{Q^N_1,\dots ,Q^N_{k_N},Q^U_1,\dots , Q^U_{k_U}\}$.
   (Intuitively, this is because state subformulas can
    be ordered according to their relative nesting.)

\- The transition function $\delta$ satisfies: for every $q \in Q$, $a \in \Sigma$
   \li
   \- if $q \in Q^N_i$, then:
      $\delta(q,a)$ contains only disjunctively related\footnotemark[1] elements of $Q^N_i$;
      every element of $\delta(q,a)$ outside of $Q^N_i$ belongs to a lower set;
   \- if $q \in Q^U_i$, then:
      $\delta(q,a)$ contains only conjunctively related\footnotemark[1] elements of $Q^U_i$;
      every element of $\delta(q,a)$ outside of $Q^U_i$ belongs to a lower set.
   \il
   \ak{figure of disj-conj related sets: and why?!}
\il
\footnotetext[1]{In a Boolean formula, atoms $E$ are disjunctively [conjunctively] related
  iff the formula can be written into DNF [CNF] in such a way that each cube [clause] has at most one element from $E$.}

Finally, $acc: Q^\omega \to \bbB$ of AHTs is defined by a set $Acc \subseteq Q$:
$acc(\pi)$ holds for $\pi=q_1q_2...\in Q^\omega$ iff one of the following holds.
\li
\- The sequence $\pi$ is trapped in some $Q^U_i$ and
   $\Inf(\pi) \cap (Acc\cap Q^U) = \emptyset$
   (co-B\"uchi acceptance).
\- The sequence $\pi$ is trapped in some $Q^N_i$ and
   $\Inf(\pi) \cap (Acc \cap Q^N) \neq \emptyset$
   (B\"uchi acceptance).
\il

An example of an alternating hesitant tree automaton is in Figure~\ref{fig:automata}.

\section{Converting \CTLstar to \LTL for Synthesis}
\label{sec:reductions-to-ltl}

\ak{
\li
\- unify: LTL vs. path formula
\- system/tree path: sync the def with its usage
\- define det/universal atm that runs on annotated computation trees?
\il
}

In this section, we describe how and why we can reduce \CTLstar synthesis
to LTL synthesis.
First, we recall the standard approach to \CTLstar synthesis,
then describe, step by step, the reduction and the correctness argument,
and then discuss some properties of the reduction.

\subsection*{LTL Encoding}

Let us first look at standard automata based algorithms for \CTLstar synthesis~\cite{informatio}\ak{find the approach in their paper}.
%,ScheweThesis}.  \sven{I am not sure if my thesis really belongs there ... .}
When synthesising a system that realizes a \CTLstar specification, we normally
\li
\- Turn the \CTLstar formula into an alternating hesitant tree automaton $A$.

\- We move from computation trees to annotated computation trees that move the (memoryless) strategy of the verifier%
   \footnote{Such a strategy maps, in each tree node, an automaton state to a next automaton state and direction.}
   into the label of the computation tree.
   This allows for using the derived universal co-B\"uchi tree automaton $U$,
   making the verifier deterministic: it does not make any decisions, as they are now encoded into the system.

\- We determinise $U$ to a deterministic tree automaton $D$.

\- We play an emptiness game for $D$.

\- If the verifier wins, his winning strategy (after projection of the additional labels) defines a system, if the spoiler wins, the specification is unrealisable.
\il

We draw from this construction and use particular properties of the alternating hesitant tree automaton $A$.
Namely, $A$ is not a general alternating tree automaton,
but is an alternating hesitant tree automaton.
Such an automaton is built from a mix of nondeterministic B\"uchi
and universal co-B\"uchi word automata,
called ``existential word automata'' and ``universal word automata''.
These universal and existential word automata start at any system state [tree node] where a universally and existentially, respectively, quantified subformula is marked as true in the annotated model [annotated computation tree].
We use the term ``existential word automata'' to emphasise that the automaton is not only a non-deterministic word automaton, but it is also used in the alternating tree automaton in a way, where the verifier can pick the system [tree] path along which it has to accept.

\begin{example}[Word and tree automata]
Consider formula $\EG\EX(g \land \X(g \land \F\neg g))$
where the propositions consist of the single output $g$ and the single input $r$.
Figure~\ref{fig:automata} shows non-deterministic word automata for the subformulas,
and the alternating (actually, nondeterministic) tree automaton for the whole formula.
In what follows, we work mostly with word automata.

\begin{figure}[tb]
\begin{subfigure}[t]{\linewidth}\center
\begin{tikzpicture}[->,>=stealth',shorten >=1pt,auto,node distance=0.94cm]
  \tikzset{every state/.style={minimum size=3mm,inner sep=0.2mm}, initial text={}}
  \tikzstyle{every edge} = [align=center,draw=black]

  \node[state,initial] (0) {$q_0$};
  \node[state] (1) [right of=0] {$q_1$};
  \node[state] (2) [right of=1] {$q_2$};
  \node[state] (3) [right of=2] {$q_3$};
  \node[state,double] (4) [right of=3] {$q_4$};

  \path
  (0) edge node {$1$} (1)
  (1) edge node {$g$} (2)
  (2) edge node {$g$} (3)
  (3) edge [loop above] node {$g$} (3)
  (3) edge node {$\neg g$} (4)
  (4) edge [loop above] node {$1$} (4);
\end{tikzpicture}
\caption{%
  NBW for $\X (g \land \X (g \land \F\neg g))$,
  the alphabet $\Sigma$ is $\{r,g\}$.
  Transitions to the non-final state $sink$ are not shown.}
\label{fig:nbw-x}
\end{subfigure}
\begin{subfigure}[t]{\linewidth}\center
\begin{tikzpicture}[->,>=stealth',shorten >=1pt,auto,node distance=0.94cm]
  \tikzset{every state/.style={minimum size=3mm,inner sep=0.2mm}, initial text={}}
  \tikzstyle{every edge} = [align=center,draw=black]

  \node[state,initial,double] (0) {$q_0'$};

  \path (0) edge [loop above] node {$p_{\EX}$} (0);
\end{tikzpicture}
\caption{%
  NBW for $\G(p_{\EX})$,
  the alphabet $\Sigma$ is $\{r, g, p_{\EX}\}$.
  The transition to the non-final state $sink$ is omitted.}
\label{fig:nbw-g}
\end{subfigure}
\begin{subfigure}[t]{\linewidth}\center
\input{figures/alternating-automaton.tikz}
\caption{Alternating tree automaton for $\EG\EX(g \land \X(g \land \F\neg g))$.
  The green color of the states indicate that they are from
  the nondeterministic partition of the states
  (and thus double-circled states are from the B\"uchi acceptance condition).
  The edges starting in the filled triangle are connected with $\land$.
  Edge label $\E$ abbreviates the set of edges, for each tree direction, connected with $\lor$.
  Thus, the transition from $q'_0$ is $((q'_0, r) \lor (q'_0,\neg r)) \land ((q_1, r) \lor (q_1,\neg r))$.
  To get an alternating automaton for $\AG\EX(...)$,
  replace in the self-loop edge of $q_0'$ label $\E$ with $\A$,
  and make the state non-final
  (these also move the state into the universal partition of the states).}
\end{subfigure}
\caption{Word and tree automata.}
\label{fig:automata}
\end{figure}
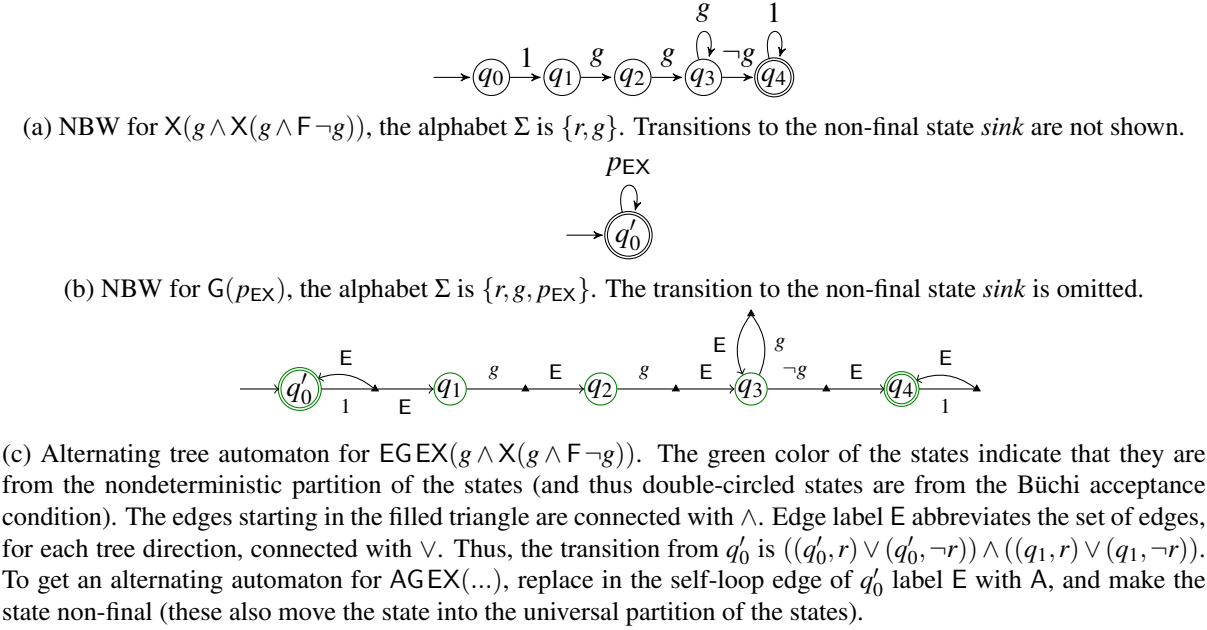
\end{example}

\smallskip

\begin{figure}[tbp]
\begin{subfigure}{\linewidth}\center
\includegraphics[width=\textwidth]{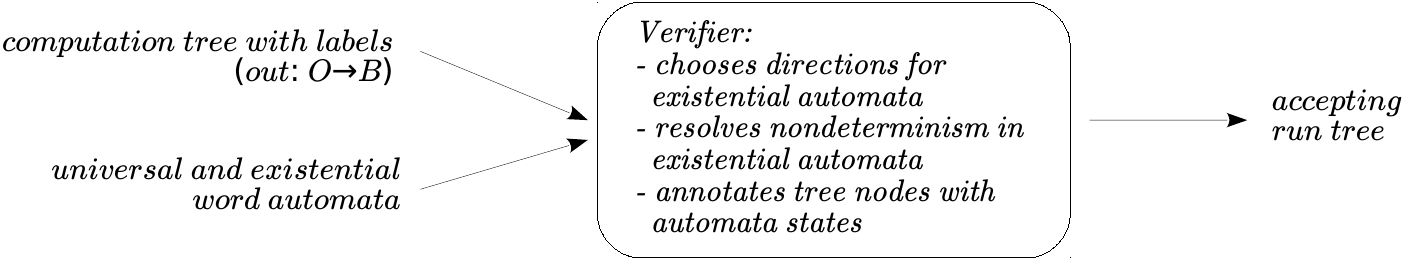}
\caption{The verifier takes a computation tree, universal and existential word automata, and the top-level proposition, that together encode a given \CTLstar formula. It produces an accepting run tree (if the computation tree satisfies the formula).}
\label{fig:stepA}
\end{subfigure}
\vspace{0.1cm}
\hrule
\vspace{0.2cm}
\begin{subfigure}{\linewidth}\center
\includegraphics[width=\textwidth]{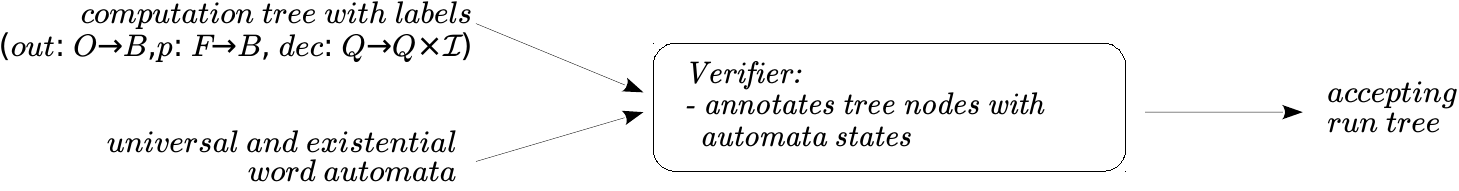}
\caption{We encode the verifier decisions into annotated computation trees,
  making the verifier deterministic.
  Figure~\ref{fig:annotated-computation-tree} shows such an annotated computation tree.}
\label{fig:stepB}
\end{subfigure}
\vspace{0.1cm}
\hrule
\vspace{0.2cm}
\begin{subfigure}{\linewidth}\center
\includegraphics[width=\textwidth]{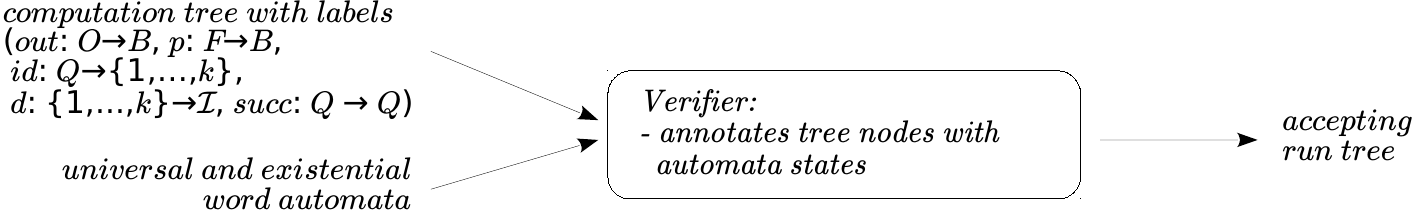}
\caption{The new annotation is a re-phrasing of the previous one.
Figure~\ref{fig:relabeled-tree} gives an example.}
\label{fig:stepC}
\end{subfigure}
\vspace{0.1cm}
\hrule
\vspace{0.2cm}
\begin{subfigure}{\linewidth}\center
\includegraphics[width=\textwidth]{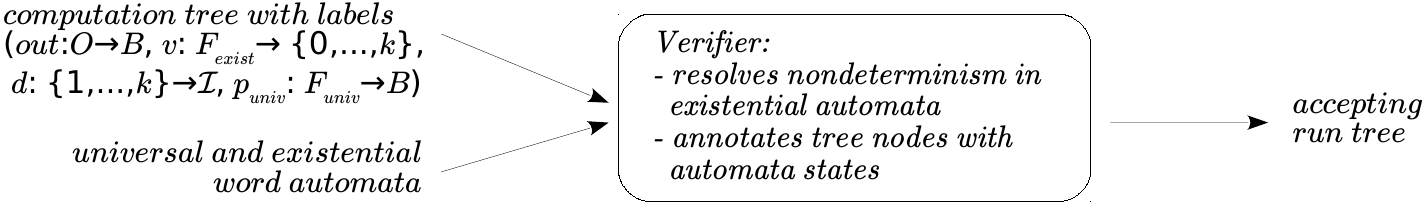}
\caption{We keep directions in the annotation but remove next-states---now the verifier has to choose.
  Figure~\ref{fig:lean-numbered-tree} gives an example.}
\label{fig:stepD}
\end{subfigure}
\vspace{0.1cm}
\hrule
\vspace{0.2cm}
\begin{subfigure}{\linewidth}\center
\includegraphics[width=\textwidth]{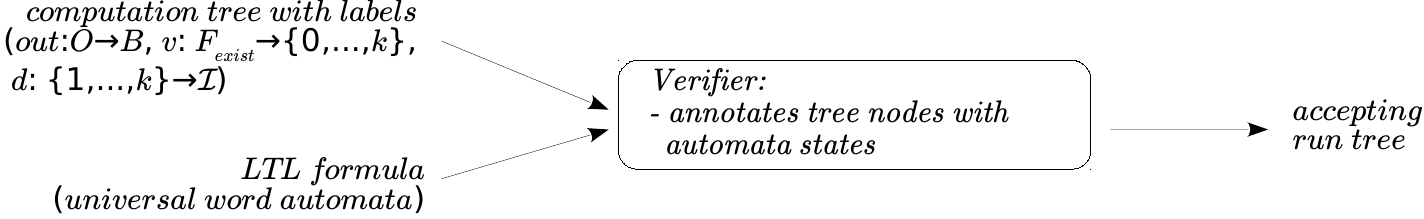}
\caption{Now the obligation of the verifier can be stated in LTL (or using universal co-B\"uchi word automata).}
\label{fig:stepE}
\end{subfigure}
\vspace{0.1cm}
\hrule
\vspace{0.2cm}
\caption{Steps in the proof of reduction of \CTLstar synthesis to LTL synthesis.}
\label{fig:discussion-summary}
\end{figure}

We are going to show, step by step,
how and why we can reduce \CTLstar-synthesis to LTL synthesis.
The steps are outlined in Figure~\ref{fig:discussion-summary}.

\parbf{Step A (the starting point)}
The verifier takes as input:
a computation tree,
universal and existential word automata for \CTLstar subformulas, and
the top-level proposition corresponding to the whole \CTLstar formula.
It has to produce an accepting run tree
(if the computation tree satisfies the formula).

\parbf{Step B}
Given a computation tree,
the verifier maps each tree node to an (universal or existential word) automaton state,
and moves from a node according to the quantification of the automaton
(either in all tree directions or in one direction).
The decision,
in which tree direction to move and which automaton state to pick for the successor node,
constitutes the strategy of the verifier.
Each time the verifier has to move in several tree directions
(this happens when the node is annotated with a \emph{universal} word automaton state),
we spawn a new version of the verifier,
for each tree direction and transition of the universal word automaton.

The strategy of the verifier is a mapping of states of the existential word automata
to a decision,
which consists of a tree direction
(the continuation of the tree path along which the automaton shall accept)
and an automaton successor state transition.
This is a mapping
$dec: Q \rightarrow \I \times Q$ \label{page:decision-mapping}
such that $dec(q)=(e,q')$ implies that $q' \in \delta\big(q,(l(n),e)\big)$,
where $\delta$ corresponds to the existential word automaton to which $q$ belongs,
and $l(n) \in \O$ is a label of the current tree node%
\footnote{%
  The verifier, when in the tree node or system state, moves according to this strategy.}%
.
Thus,
the strategy is memoryless wrt. the history of automata states\ak{note why there exists such a strategy}.
Note that strategies are defined per-node-basis,
i.e., they may be different for different nodes.

We call a model, in which every state is additionally annotated with a verifier strategy, an \emph{annotated model}.
Similarly, an \emph{annotated computation tree} is a computation tree in which every node is additionally annotated with a verifier strategy.
Thus, in both cases, every system state [node] is labeled with:
(i) original propositional labeling $out: O \to \bbB$,
(ii) propositional labeling for universal and existential subformulas $F=F_\textit{univ}\cupdot F_\textit{exist}$, $p: F \to \bbB$, and
(iii) decision labeling $dec: Q \to \I \times Q$ where $Q$ are the states of all existential automata.
\label{page:def:annotated-tree}

\ak{define when an annotated model/tree is accepted}\

\ak{state the relation btw accepted model and accepted annotated model}

\begin{example}
Figure~\ref{fig:annotated-model-tree} shows an annotated model and computation tree.
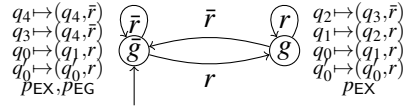
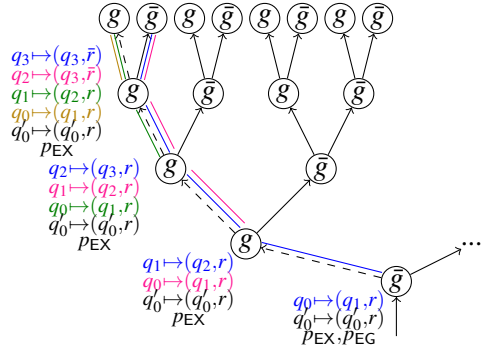
\begin{figure}[tb]
\begin{subfigure}[t]{\linewidth}\center
\input{figures/rich-model.tikz}
\caption{An annotated model satisfying $\EG\EX(g \land \X(g \land \F\neg g))$.
  Near the nodes is the winning strategy of the verifier.}
\label{fig:annotated-model}
\end{subfigure}

\vspace{0.5cm}
\begin{subfigure}[t]{\linewidth}\center
\center
\input{figures/rich-tree.tikz}
\caption{%
  An annotated computation tree that satisfies
  $\EG\EX(g \land \X(g \land \F\neg g))$.
  Let proposition $p_{\EG}$ correspond to $\EG(p_{\EX})$,
  and $p_{\EX}$---to $\EX(g \land \X(g \land \F\neg g))$.
  A winning strategy for the verifier is depicted using dashed and colored paths.
  The black dashed path witnesses $p_{\EG}$,
  the blue path witnesses $p_{\EX}$ starting in the root node,
  the pink path---$p_{\EX}$ starting in the left child, and so on.
  The pink and blue paths share the tail.
  The annotation for the verifier strategy is on the left side of nodes,
  and decisions for non mapped states are irrelevant.
  Note that this particular annotated computation tree is \emph{not} the unfolding of the annotated model above---here we postpone the right-turn of the blue path in order to illustrate that paths can share the tail.\ak{put on the right side the atm that eats such trees...and produces acc \emph{run} trees?}}
\label{fig:annotated-computation-tree}
\end{subfigure}
\caption{Annotated model and computation tree.}
\label{fig:annotated-model-tree}
\end{figure}
\end{example}

\parbf{Step C}
\ak{unify: nondet and existential atm}
The verifier strategy (encoded in the annotated computation tree) encodes both,
the words on which the nondeterministic automata are interpreted and
\hl{witnesses}\ak{define seprtly?} of acceptance
(accepting automata paths on those words).
%that this word is accepted by them.
For the encoding in LTL that we will later use,
it is enough to map out the automaton word,
%(tail of a path through the tree---or its labeling?)
%on which a nondeterministic automaton is interpreted,
and replace the witnesses by what it actually means:
that the automaton word satisfies the respective path formula.

\begin{example}
In Figure~\ref{fig:annotated-computation-tree},
the verifier strategy in the root node maps out the word
$(\bar g,p_{\EX},r) (\bar g,p_{\EX},\bar r)^\omega$
on which the NBW in Figure~\ref{fig:nbw-g} is run,
and the witness of acceptance $(q'_0)^\omega$.
The blue path encodes the word $(\bar g, r)(g,r)(g,r)(g,\bar r)(\bar g,\bar r)^\omega$
and the witness $q_0 q_1 q_2 q_3 q_3 q_4^\omega$ for the NBW in Figure~\ref{fig:nbw-x}.
In total, we can see 5 tree paths that are mapped out by the annotated computation tree.
\end{example}

To map out the word, we look at the set of tree paths
%--- and here on the tail of the paths, not their labeling ---
that are mapped out in an annotated computation tree and define equivalence classes on them.
Two tree paths are \emph{equivalent} if they share a tail
(or, equivalently, if one is the tail of the other).

There is a simple sufficient condition for two mapped out tree paths
%that are mapped out by a verifier strategy
to be equivalent:
if they pass through the same node of the annotated computation tree in the same automaton state,
then they have the same future, and are therefore equivalent.
\footnote{%
 The condition is sufficient but not necessary.
 Recall that each mapped out tree path corresponds to at least one copy of the verifier that ensures the path is accepting.
 When two verifiers go along the same tree path,
 it can be annotated with different automata states (for example, corresponding to different automata).
 Then such paths do not satisfy the sufficient condition, although they are trivially equivalent.}

\begin{example}
In Figure~\ref{fig:annotated-computation-tree} the blue and pink paths are equivalent,
since they share the tail.
The sufficient condition fires in the top node, where the tree paths meet in automaton state $q_3$
%The verifier strategy in that node could be $q_3 \mapsto (q_4,r)$ or $q_3 \mapsto (q_4,\neg r)$.
\end{example}

The sufficient condition implies that we cannot have more non-equivalent tree paths
passing through a tree node than there are states in all existential word automata; let us
call this number $k$.
%of existential automata in $A$, call this number $k$.
%Consequently, once we know (or have a lower bound for) this number, say $k$,
For each tree node, we assign unique numbers from $\{1,...,k\}$ to equivalence classes,
and thus any two non-equivalent tree paths that go through the same tree node have different numbers.
As this is an intermediate step in our translation, we are wasteful with the labeling:
\li
\-[(1)] we map existential word automata states to numbers (IDs) using a label
        $id: Q \to \{1,\ldots,k\}$,
        we choose the direction $d:\{1,\ldots,k\}\to \I$ to take,
        and choose the successor state, $succ: Q \to Q$,
        such that $succ(q) \in \delta\Big(q, \big(l(n),d(id(q))\big)\Big)$,
        where $l(n)$ is the label of the current node $n$, \emph{and}

\-[(2)] we maintain the same state ID along the chosen direction: $id(q) = id(succ(q))$.
\il

Note that (1) alone can be viewed as a re-phrasing of the labeling $dec$ that we had before on page \pageref{page:decision-mapping}.
The requirement (2) is satisfiable, because a tree path maintains its equivalence class.
Therefore any annotated computation tree can be re-labeled!\ak{better `why'!}
This step is shown in Figure~\ref{fig:stepC}, the labels are:
$(out:O\to\bbB,
p:F\to\bbB,
id: Q \to \{1,\ldots,k\},
d: \{1,\ldots,k\}\to \I,
succ: Q \to Q)$.

\begin{example}
A re-labeled computation tree is in Figure~\ref{fig:relabeled-tree}.
\begin{figure}[tb]
\center
\input{figures/relabeled-rich-tree.tikz}
\caption{%
  A re-labeled computation tree.
  Notation ``$q_0 \mapsto (1,q_1)$'' means
  $id(q_0) = 1$ and $succ(q_0) = q_1$, and ``$1 \mapsto r$'' means $d$ maps $1$ to $\{r\}$.
  Since the blue and pink paths are equivalent,
  the label $id$ maps the corresponding automata states in the nodes
  to the same number, $1$.
  The IDs of the green and yellow paths differ implying that they are not equivalent and hence do not share the tail (their tails cannot be seen in the figure).
}
\label{fig:relabeled-tree}
\end{figure}
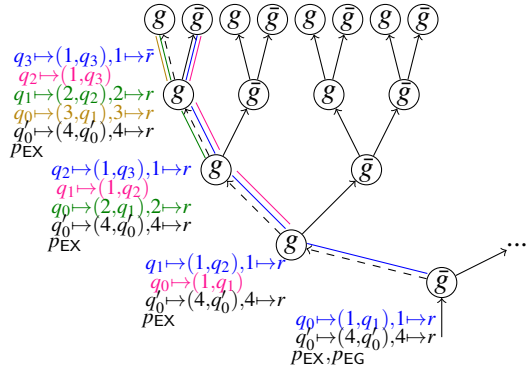
\end{example}

\parbf{Step D}
In the new annotation with labels $(out, p, id, d, succ)$,
labeling $d$ alone maps out the tree path for each ID.
The remainder of the information is mainly there to establish that the corresponding word\ak{which alphabet?}
is accepted by the respective word automaton (equivalently: satisfies the respective path formula).
%---or, likewise, that the corresponding word satisfies the respective path formula.
If we use only $d$, then the only missing information is where the path starts and which path formula it belongs to---the information originally encoded by $p$.

%We address these two points by providing this information through changing from rich to numbered computation trees: where the rich computation trees have a \emph{propositional} label for each existentially quantified path formula, we replace this propositional label by an \emph{ID}, where $0$ encodes that no claim that this subformula holds is made (similar to a proposition being ``false'' in the rich model), whereas an ID between $1$ and $k$ is interpreted like a ``true'' flag, but also requires that a respective witness is marked out with the given ID.

We address these two points by using \emph{numbered} computation trees.
Recall that the annotated computation trees have a \emph{propositional} labeling
$p: F \to \bbB$ that labels nodes with subformulas.
In the numbered computation trees,
we replace $p$ for \emph{existential} subformulas $F_\textit{exist} \subseteq F$
by labeling ${v: F_\textit{exist} \to \{0,...,k\}}$,
where, for an existentially quantified formula $\E\varphi \in F_\textit{exist}$ and a tree node $n$:
\li
\- $v_{\E\varphi}(n)=0$ encodes that no claim that $\E\varphi$ holds is made
   (similar to the proposition $p_{\E\varphi}$ being ``false'' in the annotated tree), whereas
\- a value $v_{\E\varphi}(n) \in \{1,...,k\}$ is interpreted
   \hl{similarly} to the proposition $p_{\E\varphi}$ being ``true'',
   but also requires that a witness for $\E\varphi$ is encoded on the tree path
   that starts in $n$ and follows directions $d_{v_{\E\varphi}(n)}$.
\il
% ak: "similarly" means that the verifier consults the direction-successor for the corresponding q0
% ak: "similarly" uses the fact that in in the annotated trees
% having a node marked with a subformula means that
% the node witnesses the subformula with the marked out path
% (the path is marked out by the strategy and we start in atm state q0)
% For A\phi subformulas, all paths must satisfy \phi---this is ensured
% by the verifier
% (that ensures that all such paths are accepted by the automaton).

\begin{example}
The tree in Figure~\ref{fig:relabeled-tree} becomes a numbered computation tree
if we replace the propositional labels $p_{\EX}$ and $p_{\EG}$ with ID numbers as follows.
The root node has $v_{\EX} = 1$ and $v_{\EG} = 4$,
the left child has $v_{\EX} = 1$,
the left-left child has $v_{\EX} = 2$,
the left-left-left child has $v_{\EX} = 3$.
Note that $id(q_0) = v_{\EX}$ and $id(q_0') = v_{\EG}$ whenever those $v$s are non-zero.
The nodes outside of the dashed path have $v_{\EX} = v_{\EG} = 0$,
meaning that no claims about satisfaction of the path formulas has to be witnessed there.
\end{example}

Initially, we use \emph{ID} labeling $v$ in addition with ($out, id, d, succ, p^\textit{univ}$),
where $p^\textit{univ}$ is a restriction of $p$ on $F_\textit{univ}$,
and then there is no relevant change in the way the (deterministic) verifier works.
I.e., a numbered computation tree can be turned into annotated computation tree, and vice versa,
such that the numbered tree is accepted iff the annotated tree is accepted.
% The direction rich $\leftarrow$ numbered is easy,
% the direction rich $\rightarrow$ numbered could be better formalized (?): it follows from the fact the number of equivalence classes passing through a node is less or equal to $k$.
%(Indeed: it suffices that an initial state of the respective existential word automaton is in the pre-image of ID for $id$ whenever ID is not $0$.)

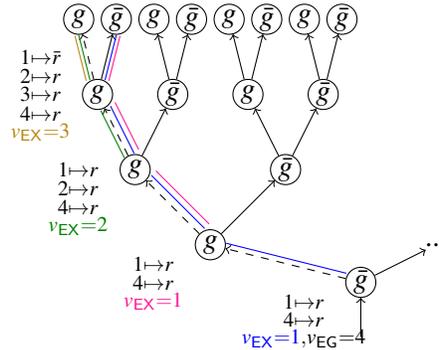
\begin{figure}[tb]
\center
\input{figures/lean-numbered-tree.tikz}
\caption{%
  Numbered computation tree with redundant annotations removed.}
\label{fig:lean-numbered-tree}
\end{figure}

\ak{define 'tree path satisfies LTL', remove 'word' in many places}

Now we observe that labeling $id$ and $succ$
are used only to witness that each word mapped out by $d$
is accepted by respective existential word automata.
I.e., $id$ and $succ$ make the verifier deterministic.
Let us remove $id$ and $succ$ from the labeling.
We call such trees \emph{lean-numbered computation trees};
they have labeling
$
(
out:O \to \bbB, 
v:F_\textit{exist} \to \{0,...,k\},
d:\{1,...,k\} \to \I,
p^\textit{univ}: F_\textit{univ} \to \bbB
)
$.
This makes the verifier nondeterministic.
We still have the property:
every accepting annotated computation tree can be turned into
an accepting lean-numbered computation tree,
and vice versa.
This step is shown in Figure~\ref{fig:stepD};
an example of a lean-numbered computation tree is in Figure~\ref{fig:lean-numbered-tree}.

\parbf{Step E (the final step)}
We show how labeling $(out,v,d,p^\textit{univ})$ 
allows for using LTL formulas instead of directly using automata for the acceptance check.
The encoding into LTL is as follows.
\li
\- For each existentially quantified formula $\E\varphi$,
   we introduce the following LTL formula
   (recall that $v_{\E\varphi} = 0$ encodes that we do \emph{not} claim that $\E\varphi$ holds in the current     tree node,
    and $v_{\E\varphi} \neq 0$ means that $\E\varphi$ does hold
    and $\varphi$ holds if we follow $v_{\E\varphi}$-numbered directions):
   \begin{equation}\label{eq:ltl-existential}
   \bigwedge_{j \in \{1,...,k\}} \G\Big[ v_{\E\varphi} = j ~\impl~ \big( \G d_j \impl \varphi'\big) \Big],
   \end{equation}
   where $\varphi'$ is obtained from $\varphi$
   by replacing the subformulas of the form $\E\psi$ by $v_{\E\psi} \neq 0$
   and the subformulas of the form $\A\psi$ by $p_{\A\psi}$.

\- For each subformula of the form $\A\varphi$, we simply take
   \begin{equation}\label{eq:ltl-universal}
   \G\Big[ p_{\A\varphi} ~\impl~ \varphi' \Big],
   \end{equation}
   where $\varphi'$ is obtained from $\varphi$ as before.

\- Finally, the overall LTL formula is the conjunction
   \begin{equation}\label{eq:ltl-full}
   \boxed{
   \Phi' \land \bigwedge_{\E\varphi \in F_\textit{exist}} \text{Eq.}\ref{eq:ltl-existential} ~\land \bigwedge_{\A\varphi \in F_\textit{univ}} \text{Eq.}\ref{eq:ltl-universal}
   }
   \end{equation}
   where the Boolean formula $\Phi'$ is obtained by replacing in the original \CTLstar formula
   every $\E\varphi$ by $v_{\E\varphi} \neq 0$ and every $\A\varphi$ by $p_{\A\varphi}$.
\il

\begin{example}\label{ex:ctlstar}
Let $I = \{r\}$, $O=\{g\}$.
Consider the \CTL formula
$$
\EG \neg g \land \AG\EF \neg g \land \EF g.
$$
The sum of states of individual NBWs is $5$
(assuming the natural translations),
so we introduce integer propositions $v_{\EF\!\bar{g}}$, $v_{\EG\!\bar{g}}$, $v_{\EF\!g}$, ranging over $\{0,...,5\}$,
%\footnote{%
%  We can slightly optimize the LTL formula by using smaller domains for $v_i$s:
%  each $v_i$ can vary over \hl{xx:non-intersecting}
%  where $Q_i$ is the number of states in the corresponding NBW.
%  }
and five Boolean propositions $d_1$, ..., $d_5$;
we also introduce Boolean proposition $p_{\AG(v_{\EF\!\bar{g}}\neq 0)}$.
The LTL formula is:
\begin{align*}
&~~~v_{\EG\!\bar{g}} \neq 0 \land p_{\AG(v_{\EF\!\bar{g}}\neq 0)} \land v_{\EF\!g} \neq 0 ~\land \\
&~~~\G\left[p_{\AG(v_{\EF\!\bar{g}}\neq 0)} \impl \G(v_{\EF\!\bar{g}}\neq 0)\right] ~\land \\
&\bigwedge_{j \in \{1...5\}}\G \left[
\begin{aligned}
& v_{\EF\!\bar{g}}=j ~\impl~ (\G d_j \impl \F \neg g) \\
& v_{\EG\!\bar{g}}=j ~\impl~ (\G d_j \impl \G \neg g) \\
& v_{\EF\!g}=j ~\impl~ (\G d_j \impl \F g)
\end{aligned}
\right]
\end{align*}
Figure~\ref{fig:ctlstar:system} shows a model satisfying the LTL specification.

\begin{figure}[bt]
\center
\begin{tikzpicture}[->,>=stealth',shorten >=1pt,auto,node distance=3.2cm]
  \tikzset{every state/.style={minimum size=6mm,inner sep=0.0mm}, initial text={}}
  \tikzstyle{every edge} = [align=center,draw=black]

  \node[state,initial below] (0)
  [label={left:\specialcellC{$v_{\EF\!\bar{g}}=v_{\EG\!\bar{g}}=2,d_2=\neg r$\\$v_{\EF\!g}=3, d_3=r$}},
   label={below right:$t_0$}] {$\neg g$};
  \node[state] (1)
  [right of=0,
  label={right:\specialcellL{$v_{\EF\!\bar{g}}=2,d_2=r$\\$v_{\EG\!\bar{g}}=v_{\EF\!g}=0$}},
  label={below left:$t_1$}] {$g$};

  \path
  (0) edge [bend left=10] node {$r$} (1)
  (1) edge [bend left=10] node {$r$} (0)
  (1) edge [loop above] node {$\neg r$} (1)
  (0) edge [loop above] node {$\neg r$} (0);
\end{tikzpicture}
\caption{A Moore machine for Example~\ref{ex:ctlstar}.
The witness for $\EG\neg g$ is:
$v_{\EG\!\bar{g}}(t_0)=2$, we move along $d_2=\neg r$ looping in $t_0$, thus the witness is $(t_0)^\omega$.
The witness for $\EF g$:
since $v_{\EF\!g}(t_0)=3$, we move along $d_3=r$ from $t_0$ to $t_1$,
where $d_3$ is not restricted, so let $d_3=\neg r$ and then the witness is $t_0 (t_1)^\omega$.
The satisfaction of $\AGEF \neg g$ means that every state has $v_{\EF\!\bar{g}} \neq 0$, which is true.
In $t_0$ we have $\neg g$, so $\EF \neg g$ is satisfied;
for $t_1$ we have $v_{\EF\!\bar{g}}(t_1)=2$ hence we move $t_1 \trans{r} t_0$ and $\EF \neg g$ is also satisfied.
}
\label{fig:ctlstar:system}
\end{figure}
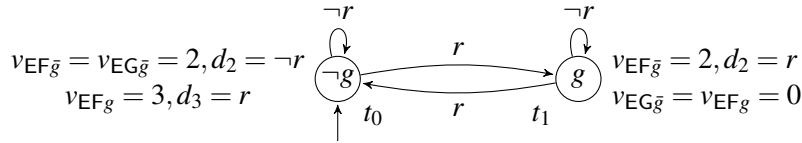
\end{example}

\medskip

%So far we have talked about computation trees.
%Finally, note that, whenever there is a computation tree satisfying an LTL formula,
%there is a finite system satisfying it~\cite{LTL-finite-model-property}.
\ak{restore this note}

Note that we can avoid introducing propositions for universally quantified subformulas
$F_\textit{univ}$:
whenever you see such a proposition
in $\varphi'$ in Eq.~\ref{eq:ltl-existential}
or in $\Phi'$ in Eq.~\ref{eq:ltl-full},
replace it with subformula $\varphi''$ which it describes.

\smallskip

The whole discussion leads us to the theorem.
\begin{theorem}
  Let $I$ be the set of inputs and $O$ be the set of outputs,
  and $\Phi_\LTL$ be derived from a given $\Phi_\CTLstar$ as described above.
  Then:
  $$
  \textit{
  $\Phi_\CTLstar$ is realisable
  $~\Iff~$
  $\Phi_\LTL$ is realisable.
  }
  $$
\end{theorem}

\subsection*{Complexity}
The translated LTL formula $\Phi_\text{LTL}$, due to Eq.~\ref{eq:ltl-existential},
in the worst case, can be exponentially larger than $\Phi_\CTLstar$,
$|\Phi_\LTL| = 2^{\Theta(|\Phi_\CTLstar|)}$.
Yet, the upper bound on the size of $UCW_{\Phi_\LTL}$ is $2^{\Theta(|\Phi_\CTLstar|)}$
rather than $2^{\Theta(|\Phi_\LTL|)}=2^{2^{\Theta(|\Phi_\CTLstar|)}}$,
because:
\li
\- the size of the UCW is additive in the size of the UCWs of the individual conjuncts, and
\- each conjunct UCW has almost the same size as a UCW of the corresponding subformula,
   since, for every LTL formula $\varphi$, $|UCW_{\G[p \impl (\G\!d \impl \varphi)]}| = |UCW_\varphi|+1$.%
   \footnote{To see this, recall that we can get $UCW_\psi$ by
     treating $NBW_{\neg\psi}$ as a UCW,
     and notice that $|NBW_{\F[p\land\G\!d\land\neg\varphi]}| = |NBW_{\neg\varphi}|+1$.}
\il
Determinising $UCW_{\Phi_\LTL}$ gives a parity game with up to $2^{2^{\Theta(|\Phi_\CTLstar|)}}$ states and
$2^{\Theta(|\Phi_\CTLstar|)}$ priorities~\cite{Schewe/09/determinise,Piterman07,Safra}.
The recent quasipolynomial algorithm~\cite{DBLP:conf/stoc/CaludeJKL017} for solving parity games
has a particular case for $n$ states and $log(n)$ many priorities,
where the time cost is polynomial in the number of game states.
This gives us $O(2^{2^{|\Phi_\CTLstar|}})$-time solution to the derived LTL synthesis problem.
The lower bound comes from the 2EXPTIME-completeness of the \CTLstar synthesis problem~\cite{RosnerThesis}.

\begin{theorem}
  Our solution to the \CTLstar synthesis problem
  via the reduction to \LTL synthesis
  is 2EXPTIME-complete.
\end{theorem}

\subsection*{Minimality}
Although the reduction to LTL synthesis preserves the complexity class,
it does not preserve the minimality of the models.
Consider an existentially quantified formula $\E\varphi$.
A system path satisfying the formula may pass through the same system state more than once
and exit it in different directions.%
  \footnote{%
  E.g., in Figure~\ref{fig:annotated-model} the system path $t_0 t_1 t_1 (t_0)^\omega$,
  satisfying $\EX(g\land\X(g\land\F\neg g))$,
  double-visits state $t_1$ and exits it first in direction $r$ and then in $\neg r$,
  where $t_0$ is the system state on the left and $t_1$ is on the right.}
Our encoding forbids that.%
  \footnote{%
  Recall that with $\E\varphi$ we associate a number $v_{\E\varphi}$,
  such that whenever in a system state $v_{\E\varphi}$ is non-zero,
  then the path mapped out by $v_{\E\varphi}$-numbered directions satisfies the path formula $\varphi$.
  Therefore whenever $v_{\E\varphi}$-numbered path visits a system state,
  it exits it in the \emph{same} direction $d_{v_{\E\varphi}}$.
  %I.e., a $v_{\E\varphi}$-numbered path can visit a \ul{system state}
  %zero times, once, or infinitely many times.
  %I.e., any $v_{\E\varphi}$-numbered path is a simple lasso.
  }
I.e., in any system satisfying the derived LTL formula,
a system path mapped out by an ID has a unique outgoing direction from every visited state.
As a consequence, such systems are less concise.\ak{what is the worst case blowup?}
This is illustrated in the following example.

\begin{example}[Non-minimality]\label{ex:ctlstar:nonminimal}
Let $I = \{r\}$, $O=\{g\}$,
and consider the \CTLstar formula
$$
\EX(g \land \X(g \land \F\neg g))
$$
The NBW automaton for the path formula has 5 states (Figure~\ref{fig:nbw-x}),
so we introduce integer proposition $v$ ranging over $\{0,...,5\}$
and Boolean propositions $d_1$, $d_2$, $d_3$, $d_4$, $d_5$.
The LTL formula is
$$
~~~v \neq 0 ~\land
\bigwedge_{j \in \{1...5\}}\!\!\!\!
\G\big[v=j ~\impl~ (\G d_j \impl \X(g \land \X(g \land \F\neg g)))\big]
$$
A smallest system for this LTL formula is in Figure~\ref{fig:ctlstar:nonminimal:system:ltl}.
It is of size is $3$,
while a smallest system for the original \CTLstar formula is of size $2$
(Figure~\ref{fig:annotated-model}).
\begin{figure}[tb]\center
\begin{tikzpicture}[->,>=stealth',shorten >=1pt,auto,node distance=2.2cm]
  \tikzset{every state/.style={minimum size=6mm,inner sep=0.0mm}, initial text={}}
  \tikzstyle{every edge} = [align=center,draw=black]

  \node[state,initial] (0)
  [label={above right:\specialcellC{$v=1$\\$d_1=\neg r$}},
   label={below:$t_0$}] {$\neg g$};

  \node[state] (1)
  [right of=0,
  label={right:{$d_1=r$}},
  label={below left:$t_1$}] {$g$};

  \node[state] (2)
  [below of=1,
  label=left:{$t_2$},
  label=right:{$d_1=r$}
  ] {$g$};

  \path
  (0) edge node[below] {$\neg r$} (1)
  (1) edge [loop above] node {$\neg r$} (1)
  (1) edge node {$r$} (2)
  (2) edge node {$1$} (0)
  (0) edge [loop above] node {$r$} (0);
\end{tikzpicture}
\caption{%
  A smallest Moore machine satisfying the LTL formula from Example~\ref{ex:ctlstar:nonminimal}.}
\label{fig:ctlstar:nonminimal:system:ltl}
\end{figure}
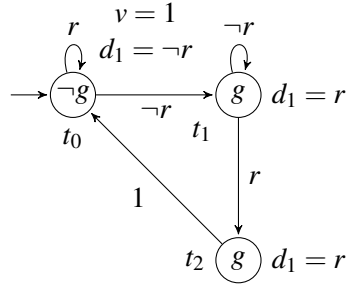
\end{example}

\subsection*{Bounded reduction}
While we have realisability equivalence for sufficiently large $k$, $k$ is a parameter,
where much smaller $k$ might suffice.
In the spirit of bounded synthesis,
it is possible to use smaller parameters in the hope of finding a model.
These models might be of interest in that they guarantee a limited entanglement of different tree paths,
as they cap the number of tails of tree paths that go through the same node of a computation tree.
Such models are therefore simple in some formal sense,
and this sense is independent of the representation by an automaton.
(As opposed to a lower bound of a sufficiently high number $k$,
 for which we have explicitly used the representation by an automaton.) 

\ak{CTL?}

\section{Checking Unrealisability of \CTLstar} \label{sec:ctlstar-unreal}
How does a witness of unrealisability for \CTLstar look like?
I.e., when a formula is unrealisable,
is there an ``environment model'', like in the LTL case,
which disproves any system model?

The LTL formula and the annotation shed light on this:
the model for the dualised case is a strategy to choose original inputs
(depending on the history of $v$, $d$, $p$, and original outputs),
such that any path in the resulting tree violates the original LTL formula.
I.e., the spoiler strategy is a tree, whose nodes are labeled with original inputs,
and whose directions are defined by $v$, $d$, $p$, and original outputs.

% The spoiler would prefer to annotate tree nodes in conformance with some direction $d$
% and to satisfy $\neg\varphi'$
% in Eq.~\ref{eq:ltl-existential} or Eq.~\ref{eq:ltl-universal},
% whenever the implications are triggered.
% I.e. the spoiler has a current direction to follow,
% as well as a set of ``promising'' directions,

\begin{example}
Consider an unrealisable \CTLstar specification:
$\AG g \land \EFX\neg g$, inputs $\{r\}$, outputs $\{g\}$.
After reduction to LTL we get specification:
inputs $\{r\}$, outputs $\{g,p_{\AG\!g},v_{\EFX\!\bar g},d_1,d_2\}$,
and the LTL formula is
$$
p_{\AG\!g} \land v_{\EFX\!\bar g} \neq 0 \land
\G\big[ p_{\AG\!g} \impl \G g \big] \land\!\!\!
\bigwedge_{j \in \{1,2\}}\!\!\!\!\!\G\big[ (v_{\EFX\!\bar g} = j \land \G d_j)  \impl  \F\!\X\neg g \big].
$$
The dual specification is:
the system type is Mealy,
new inputs $\{g,p_{\AG\!g},v_{\EFX\!\bar g},d_1,d_2\}$,
new outputs $\{r\}$,
and the LTL formula is the negated original LTL:
$$
p_{\AG\!g} \land v_{\EFX\!\bar g} \neq 0 \land
\G\big[ p_{\AG\!g} \impl \G g \big] ~\impl\!
\bigvee_{j \in \{1,2\}}\!\!\!\!\!\F\big[ (v_{\EFX\!\bar g} = j \land \G d_j)  \land  \G\!\X g \big].
$$
This dual specification is realisable, and it exhibits e.g. the following witness of unrealisability:
the output $r$ follows $d_1$ or $d_2$ depending on input $v_{\EFX\!\bar g}$.
(The new system needs two states.
 State $1$ describes ``I've seen $v_{\EFX\!\bar g}\in\{0,1\}$ and I output $r$ equal to  $d_1$'';
 from state $1$ we irrevocably go into state $2$ once $v_{\EFX\!\bar g}=2$ and make $r$ equal to $d_2$).
\end{example}

Although our encoding allows for checking unrealisability of \CTLstar
(via dualising the converted LTL specification),
this approach suffers from a very high complexity.
Recall that the LTL formula can become exponential in the size of a \CTLstar formula, which could only be handled, because it became a big conjunction with sufficiently small conjuncts.
After negating, it becomes a large disjunction,
which makes the corresponding UCW doubly exponential in the size of the initial \CTLstar specification
(vs.\ single exponential for the non-negated case).
%Not only can $k$ be exponential in the size of the $\CTLstar$ formula making the LTL formula exponential,
%all is multiplicative here. % we cannot seem to bound the game size by 2EXP (3EXP --- yes)
This seems---there may be a more clever analysis of the formula structure---%
to make the unrealisability check via reduction to LTL cost three exponents in the worst case
(vs.\ 2EXP by the standard approach).

What one could try is to let the new system player in the dualised game
choose a number of disjunctive formulas to follow,
and allow it to revoke the choice finitely many times.\ak{clarify}
This is conservative:
if following $m$ different disjuncts in the dualised formula is enough to win,
then the new system wins.
% There does not seem to be good complexity guarantees that go with this,
% but with a bit of luck that might work.
Also, parts of the disjunction might work well (``delta-debugging'');
this could then be handled precisely.

\iffinal\else{
\hl{%
  Is there a bounded procedure which is like $2^{2^{|\Phi_\CTLstar|}\cdot 2^{\textit{smallest size}}}$
  where the 2exp is in the SMT solver? (just like in the bounded synthesis?)}
}{}
\fi

\section{Experiments}\label{sec:experiments}

We implemented the \CTLstar to LTL converter {\small\tt ctl\_to\_ltl.py} inside PARTY~\cite{party}.
PARTY also has two implementations of the bounded synthesis approach~\cite{BS},
one encodes the problem into SMT and another reduces the problem to safety games.
Also, PARTY has a \CTLstar synthesiser~\cite{CTLstarCAV} based on the bounded synthesis idea
that encodes the problem into SMT.
In this section we compare those three solvers, where the first two solvers take LTL formulas produced
by our converter.
All logs and the code are available in repository {\small\url{https://github.com/5nizza/party-elli}},
the branch ``cav17''.
The results are in Table~\ref{tab:optimizations}, let us analyse them.

\begin{table}[tb]
\caption{Comparison of different synthesis approaches for \CTLstar specifications.
  All specifications are realisable.
  $|\CTLstar|$ is the size of the non-reduced AST of the \CTLstar formula,
  $|\LTL|$ --- similarly, but it has two numbers:
  when the parameter $k$ is set to $1$ ($k$ is the number of witness IDs),
  and when $k$ is the upper bound (the number of existential states).
  $|AHT|$ is the sum of the number of automata states for all subformulas.
  $|UCW|$ is the number of states in the UCW of the translated LTL formula:
  we show two numbers, when $k$ is set to $1$ and when it is the upper bound.
  Timings are in seconds, the timeout is 3 hours (denoted ``$to$'').
  ``Time \CTLstar'' is the synthesis time and [model size] required for
  \CTLstar synthesizer {\small \tt star.py},
  ``time LTL(SMT)'' --- for synthesizer {\small\tt elli.py} which implements the original bounded synthesis for LTL via SMT~\cite{BS},
  ``time LTL(game)'' --- for synthesizer {\small\tt kid.py} which implements the original bounded synthesis for LTL via reduction to safety games~\cite{BS}.
  Both ``time LTL'' columns have two numbers:
  when $k$ is set to the minimal value for which the LTL is realisable,
  and when $k$ is set to the upper bound.
  The subscript near the number indicates the value of $k$:
  e.g. $to_8$ means the timeout on all values of $k$ from 1 to $|Q|=8$;
  $to_{12(3)}$ means there was the timeout for $k=|Q|=12$
  and the last non-timeout was for $k=3$;
  $20_1$ means 20 seconds and the minimal $k$ is 1.
  The running commands were:
    ``{\small \tt elli.py --incr spec}'',
    ``{\small \tt star.py --incr spec}'',
    ``{\small \tt kid.py spec}''.
  }
\scriptsize
\centering
\setlength{\tabcolsep}{3pt}
\begin{tabular}{ lcc|cc|r|cc }
\toprule
  & $|\CTLstar|$
  & \specialcellC{$|\LTL|$\\($k_{1}$:$k_{|Q|}$)}
  & $|\text{AHT}|$
  & \specialcellC{$|\text{UCW}|$\\($k_{1}$:$k_{|Q|}$)}
  & \specialcellC{time\\ \CTLstar} & \specialcellC{time\\LTL(SMT)\\($k_{min}$:$k_{|Q|}$)} & \specialcellC{time\\LTL(game)\\($k_{min}$:$k_{|Q|}$)} \\
\midrule
res\_arbiter3     &  65   &  ~78 : 127  &  9   &   7 : 9    & 25 [5]         &  $40_1:260_2$              &  $\mathbf{~7_1:20_2}$   \\
res\_arbiter4     &  97   &  109 : 168  &  10  &  ~8 : 10   & 7380 [7]       &    $to_1$                  &  $\mathbf{30_1:60_2}$   \\
loop\_arbiter2    &  49   &  105 : 682  &  12  &  11 : 41   & {\bf 2} [4]    &  $20_3:131_6$              & ~~~$18_3:to_{6(5)}$   \\
loop\_arbiter3    &  80   & ~183 : 1607 &  15  &  14 : 70   & {\bf 6360} [7] &  $to_{8}$                  & $to_{8}$   \\
postp\_arbiter3   &  113  & ~177 : 2097 &  19  & ~~15 : 114 & 3 [4]          &  ~~~~$\mathbf{2_1}:1735_{12}$  & ~~~~$20_1:to_{12(3)}$   \\
postp\_arbiter4   &  162  & ~276 : 4484 &  24  &  19 : $to$ & 2920 [5]       &  ~~~$\mathbf{68_1}:to_{16(5)}$  & ~~~~$70_1:to_{16(2)}$  \\
prio\_arbiter2    &  82   &  ~92 : 141  &  13  &  14 : 16   & 60 [5]         &    $14_1:19_2$~            & ~$\mathbf{9_1:17_2}$   \\
prio\_arbiter3    &  117  &  125 : 184  &  15  &  16 : 18   & $to$~~~~       &  $4318_1:to_2$~~~~         & ~$\mathbf{26_1:56_2}$~   \\
user\_arbiter1    &  99   & ~190 : 4385 &  23  &  25 : $to$ &  {\bf 3} [5]   &  $1855_5:to_{16}$~~~       & $to_{16}$  \\
%\midrule
\bottomrule
\end{tabular}
\label{tab:optimizations} 
\end{table}

\parbf{Specifications}
We created realisable arbiter-like \CTLstar specifications.
The number after the specification name indicates the number of clients.
All specifications have LTL properties in the spirit of ``every request must eventually be granted'' and the mutual exclusion of the grants.
Also:
\li
\- ``res\_arbiter'' has the resettability property $\AG\EFG(\bigwedge_i\!\neg g_i)$;
\- ``loop\_arbiter'' in addition has the looping property $\bigwedge_i\!\EFG g_i$;
\- ``postp\_arbiter'' has the \CTLstar property $\bigwedge_i\AGEF(\neg g_i \land r_i \land \X(\neg g_i \land r_i \land X \neg g_i ))$;
\- ``prio\_arbiter'' prioritizes requests from one client (this is expressed in LTL), and has the resettability property;
\- ``user\_arbiter'' contains only existential properties
   that specify different sequences of requests and grants.
\il

\parbf{LTL formula and automata sizes}
LTL formula increases $\approx |Q|$ times when $k$ increases from $1$ to $|Q|$,
just as described by Eq.~\ref{eq:ltl-existential}.
But this increase does not incur the exponential blow up of the UCWs:
they also increase only $\approx |Q|$ times.

\parbf{Synthesis time}
The game-based LTL synthesiser is the fastest in half of the cases,
but struggles to find a model when $k$ is large.
The LTL part of specifications ``res\_arbiter'' and ``prio\_arbiter'' is known to be simpler
for game-based synthesisers than for SMT-based ones---adding the simple resettability property does not change this.

\parbf{Model sizes}
The reduction did not increase the model size in all the cases.

\section{Conclusion}\label{sec:conclusion}
We presented the reduction of \CTLstar synthesis problem to LTL synthesis problem.
The reduction preserves the worst-case complexity of the synthesis problem,
although possibly at the cost of larger systems.
The reduction allows the designer to write \CTLstar specifications
even when she has only an LTL synthesiser at hand.
We experimentally showed---on the \emph{small} set of specifications---%
that the reduction is practical when the number of existentially quantified formulas is small.

We briefly discussed how to handle unrealisable \CTLstar specifications.
Whether our suggestions are practical on typical specifications---%
this is still an open question.
A possible future direction is to develop a similar reduction for logics like ATL*~\cite{Alur97},
and to look into the problem of satisfiability of \CTLstar~\cite{ES84}.

\ack{This work
  was supported by the Austrian Science Fund (FWF) under the RiSE
  National Research Network (S11406), and by the EPSRC through grant
  EP/M027287/1 (Energy Efficient Control).
  We thank SYNT organisers for providing the opportunity to improve the paper,
  and reviewers for their patience.}

\bibliographystyle{eptcs}
\bibliography{refs}
\end{document}

%% file: macros.tex
\usepackage{amsthm}
\usepackage{amsfonts}
\usepackage{amsmath}
\usepackage{amssymb}
\usepackage{temporal}
\usepackage[dvipdfmx]{graphicx} 
\usepackage{caption}
\usepackage{xspace}
\usepackage{subcaption}
\usepackage{stmaryrd} % for short arrows
\usepackage{multirow}
\usepackage{underscore}   % to allow _ in doins

\usepackage[svgnames]{xcolor}  % used by preamble
\usepackage{wrapfig}
\usepackage{booktabs}  % better tables
\usepackage{colortbl}
\usepackage{float}  % for forcing position with 'H' (like \begin{table}[H])

%\pdfsuppresswarningpagegroup=1  % to silence Latex warnings about multiple PDFs included in a single page

%%%%%%%%%%%%%% trackchanges and related %%%%%%%%%%%%%%
% ignore llncs note command
\let\note\relax

\iffinal
  \usepackage[finalold]{trackchanges} % also use \sr,\sj,\ak,\ba
\else
  \usepackage[inline]{trackchanges}
\fi

\addeditor{AK}
\addeditor{RB}
\addeditor{Sven}
\newcommand{\ak}[1]{\note[AK]{#1}}
\newcommand{\rb}[1]{\note[RB]{#1}}

\definecolor{lightgray}{gray}{0.8}
\definecolor{darkgreen}{rgb}{0,0.5,0}
\definecolor{darkblue}{rgb}{0,0,.5}
\definecolor{darkred}{rgb}{0.9,0,0}
\definecolor{pink}{rgb}{0.95,0.08,0.55}
\definecolor{mygray}{gray}{.6}

%% end of trackchanges

% goes last
\usepackage{hyperref}
\hypersetup{
    colorlinks,
    linkcolor=black,
    citecolor=black,
}

%---------------------------------------------------------------------
%---------------------------------------------------------------------
%---------------------------------------------------------------------
%---------------------------------------------------------------------
%---------------------------------------------------------------------
%---------------------------------------------------------------------

\newcommand{\mailto}[1]{\href{mailto:#1}{\nolinkurl{#1}}}
\newcommand{\ack}[1]{\smallskip\noindent\small{\textbf{Acknowledgements.} #1}}

\newtheoremstyle{myexample}% name
  {5pt}%      Space above
  {5pt}%      Space below
  {}%         Body font
  {}%         Indent amount (empty = no indent, \parindent = para indent)
  {\itshape}% Thm head font
  {.}%        Punctuation after thm head
  {.5em}%     Space after thm head: " " = normal interword space;
        %       \newline = linebreak
  {}%         Thm head spec (can be left empty, meaning `normal')

{\theoremstyle{remark} \newtheorem{remark}{Remark} }
{\theoremstyle{myexample} \newtheorem{example}{Example} }
\newtheorem{theorem}{Theorem}
%\declaretheorem[name=Theorem]{thm}
%\declaretheorem[name=Lemma]{lem}
%\declaretheorem[name=Corollary]{cor}
%\declaretheorem[name=Observation]{obs}
%\declaretheorem[name=Claim]{claim}

% Formatting: paragraph (bold and italic)
\newcommand{\parbf}[1]{\smallskip\noindent {\bf #1.}}

% Formatting enumerations:
\newcommand\li{\begin{itemize}}
\newcommand\il{\end{itemize}}
\renewcommand{\-}{\item}
\newcommand\lo{\begin{enumerate}}
\newcommand\ol{\end{enumerate}}

% math
\newcommand{\bbN}{\mathbb{N}}
\newcommand{\bbB}{\mathbb{B}}

\renewcommand{\|}{\ensuremath{\mid}} %% 'such that'

\newcommand{\cupdot}{\mathbin{\dot{\cup}}} %% disjoint union

\newcommand{\Iff}{\Leftrightarrow}
\newcommand{\impl}{\rightarrow}

\newcommand{\trans}[1]{\stackrel{{#1}}{\rightarrow}}

\renewcommand{\true}{\textsf{true}\xspace}
\renewcommand{\false}{\textsf{false}\xspace}

% add line breaks to table cells, to labels of tikz 
% usage: \specialcellC{first line \\ second line} 
\newcommand{\specialcellC}[2][c]{%
  \begin{tabular}[#1]{@{}c@{}}#2\end{tabular}}
\newcommand{\specialcellL}[2][c]{%
  \begin{tabular}[#1]{@{}l@{}}#2\end{tabular}}

% temporal logic
\newcommand\G\always\xspace
\newcommand\F\eventually\xspace
\newcommand\W\weakuntil\xspace
\newcommand\U\until\xspace
\newcommand\R\releases\xspace
\newcommand\X\nextt\xspace

\newcommand\E{{\ensuremath\pexists}\xspace}
\newcommand\A{{\ensuremath\pforall}\xspace}

% ttfamily with bold. Use in lstslistings: 
% \begin{lstlisting}[basicstyle=\ttfamilywithbold,language=python,mathescape]
%   the code here will have bold
% \end{lstlisting}
% http://tex.stackexchange.com/questions/25249/how-do-i-use-a-particular-font-for-a-small-section-of-text-in-my-document/25251#25251
%\newcommand*{\ttfamilywithbold}{\fontfamily{lmtt}\selectfont}

% other math macros

%\newcommand{\I}{{\cal I}}
\newcommand{\I}{{2^I}}
\renewcommand{\O}{{2^O}}

\newcommand\LTL{\ensuremath{\text{LTL}}\xspace}
\newcommand\CTLstar{\ensuremath{{\text{CTL}^*}}\xspace}
\newcommand\CTL{\ensuremath{\text{CTL}}\xspace}

\newcommand{\Inf}{\textit{Inf}}

\iffinal
\renewcommand{\hl}[1]{#1}
\else
\fi

%%%%%%%%%%%%%%%%%%%%%%%%%%%%%%%%%%%%%%%%%%%%%%%%%%%%%%%%%%%%
% preamble used for pictures by tikzit
% I don't know how to import the preamble though:
% of course, you can just copy-paste it into the custom preamble,
% but the default preamble should also be changed so that tikzit can correctly display it
%    but tikzit does not allow to modify the default preamble nor to import it...

%\usepackage[svgnames]{xcolor}
\usepackage{tikz}
\usetikzlibrary{arrows,automata}
\usetikzlibrary{decorations.markings}
\usetikzlibrary{shapes.geometric}

\pgfdeclarelayer{edgelayer}
\pgfdeclarelayer{nodelayer}
\pgfsetlayers{edgelayer,nodelayer,main}

\tikzstyle{none}=[inner sep=0pt]
\tikzset{initial text={}}

\tikzstyle{rn}=[circle,fill=White,draw=Red,minimum size=0.4cm,inner sep=0pt]
\tikzstyle{gn}=[circle,fill=White,draw=Green,minimum size=0.4cm,inner sep=0pt]
\tikzstyle{yn}=[circle,fill=White,draw=Yellow,minimum size=0.4cm,inner sep=0pt]
\tikzstyle{wn}=[circle,fill=White,draw=Black,minimum size=0.4cm,inner sep=0pt]
\tikzstyle{invisible}=[circle,fill=White,draw=White,inner sep=0pt]
\tikzstyle{textual}=[rectangle]
\tikzstyle{dot}=[circle,fill=Black,draw=Black,inner sep=0pt,minimum size=0.5mm]
\tikzstyle{uptriangle}=[regular polygon,regular polygon sides=3,shape border rotate=0,fill=Black,draw=Black,inner sep=0pt,minimum size=1mm]

\tikzstyle{simple}=[-,draw=Black]
\tikzstyle{arrow}=[->,draw=Black]
\tikzstyle{dashed arrow}=[->,draw=Black,dashed]
\tikzstyle{tick}=[-,draw=Black,postaction={decorate},decoration={markings,mark=at position .5 with {\draw (0,-0.1) -- (0,0.1);}}]
\tikzstyle{ga}=[->,draw=Green]
\tikzstyle{pa}=[->,draw=DeepPink]
\tikzstyle{red}=[-,draw=Red]
\tikzstyle{pink}=[-,draw=DeepPink]
\tikzstyle{blue}=[-,draw=Blue]
\tikzstyle{green}=[-,draw=Green]
\tikzstyle{yellow}=[-,draw=DarkGoldenrod]

%%%% end of tikzit preamble
%%%%%%%%%%%%%%%%%%%%%%%%%%%%%%%%%%%%%%%%%%%%%%%%%%%%%%%%%%%%

%% file: figures/alternating-automaton.tikz
\begin{tikzpicture}
	\begin{pgfonlayer}{nodelayer}
		\node [style=gn, initial, double] (0) at (0, -0) {$q_0'$};
		\node [style=gn] (1) at (2, -0) {$q_1$};
		\node [style=uptriangle] (2) at (1, -0) {};
		\node [style=gn] (3) at (4, -0) {$q_2$};
		\node [style=gn] (4) at (6, -0) {$q_3$};
		\node [style=gn, double] (5) at (8, -0) {$q_4$};
		\node [style=uptriangle] (6) at (7, -0) {};
		\node [style=uptriangle] (7) at (5, -0) {};
		\node [style=uptriangle] (8) at (3, -0) {};
		\node [style=uptriangle] (9) at (9, -0) {};
		\node [style=uptriangle] (10) at (6, 1) {};
	\end{pgfonlayer}
	\begin{pgfonlayer}{edgelayer}
		\draw [style=simple] (0) to node[below]{$_1$} (2);
		\draw [style=arrow] (2) to node[below=]{$_{\sf E}$} (1);
		\draw [style=arrow, bend right, looseness=1.00] (2) to node[above]{$_{\sf E}$} (0);
		\draw [style=simple] (1) to node[above]{$_g$} (8);
		\draw [style=simple] (3) to node[above]{$_g$} (7);
		\draw [style=simple] (4) to node[above]{$_{\neg g}$} (6);
		\draw [style=simple] (5) to node[below]{$_1$} (9);
		\draw [style=arrow] (8) to node[above]{$_{\sf E}$} (3);
		\draw [style=arrow] (7) to node[above]{$_{\sf E}$} (4);
		\draw [style=arrow] (6) to node[above]{$_{\sf E}$} (5);
		\draw [style=arrow, bend right, looseness=1.00] (9) to node[above]{$_{\sf E}$} (5);
		\draw [style=simple, bend right, looseness=1.00] (4) to node[right]{$_g$} (10);
		\draw [style=arrow, bend right, looseness=1.00] (10) to node[left]{$_{\sf E}$} (4);
	\end{pgfonlayer}
\end{tikzpicture}

%% file: figures/rich-model.tikz
\begin{tikzpicture}
	\begin{pgfonlayer}{nodelayer}
		\node [style=wn, initial below] (0) at (0, -0) {$\bar g$};
		\node [style=wn] (1) at (2, -0) {$g$};
		\node [style=textual] (2) at (-1, -0) {$\color{black}{_{q_0 \mapsto (q_1,r)}}$};
		\node [style=textual] (3) at (-1, -0.25) {$_{q_0' \mapsto (q_0',r)}$};
		\node [style=textual] (4) at (3, -0) {$\color{black}{_{q_0 \mapsto (q_1,r)}}$};
		\node [style=textual] (5) at (3, -0.25) {$_{q_0' \mapsto (q_0',r)}$};
		\node [style=textual] (6) at (-1, 0.25) {$\color{black}{_{q_3 \mapsto (q_4,\bar r)}}$};
		\node [style=textual] (7) at (3, 0.25) {$\color{black}{_{q_1 \mapsto (q_2,r)}}$};
		\node [style=textual] (8) at (3, 0.5) {$\color{black}{_{q_2 \mapsto (q_3,\bar r)}}$};
		\node [style=textual] (9) at (-1, 0.5) {$\color{black}{_{q_4 \mapsto (q_4,\bar r)}}$};
		\node [style=textual] (10) at (3, -0.5) {$_{p_{\sf EX}}$};
		\node [style=textual] (11) at (-1, -0.5) {$_{p_{\sf EX}, p_{\sf EG}}$};
	\end{pgfonlayer}
	\begin{pgfonlayer}{edgelayer}
		\draw [style=arrow, bend right=15, looseness=1.00] (0) to node[below]{$r$} (1);
		\draw [style=arrow, bend right=15, looseness=1.00] (1) to node[above]{$\bar r$} (0);
		\draw [style=arrow, in=120, out=60, loop] (0) to node[below]{$\bar r$} ();
		\draw [style=arrow, in=120, out=60, loop] (1) to node[below]{$r$} ();
	\end{pgfonlayer}
\end{tikzpicture}

%% file: figures/rich-tree.tikz
\begin{tikzpicture}
	\begin{pgfonlayer}{nodelayer}
		\node [style=invisible] (0) at (1, 1) {...};
		\node [initial below, style=wn] (1) at (0, 0.5) {$\bar g$};
		\node [style=wn] (2) at (-2, 1) {$g$};
		\node [style=wn] (3) at (-3, 2) {$g$};
		\node [style=wn] (4) at (-1, 2) {$\bar g$};
		\node [style=wn] (5) at (-0.5, 3) {$\bar g$};
		\node [style=wn] (6) at (-1.5, 3) {$g$};
		\node [style=wn] (7) at (-2.5, 3) {$\bar g$};
		\node [style=wn] (8) at (-3.5, 3) {$g$};
		\node [style=wn] (9) at (-3.75, 4) {$g$};
		\node [style=wn] (10) at (-3.25, 4) {$\bar g$};
		\node [style=wn] (11) at (-2.75, 4) {$g$};
		\node [style=wn] (12) at (-2.25, 4) {$\bar g$};
		\node [style=wn] (13) at (-1.75, 4) {$g$};
		\node [style=wn] (14) at (-1.25, 4) {$\bar g$};
		\node [style=wn] (15) at (-0.75, 4) {$g$};
		\node [style=wn] (16) at (-0.25, 4) {$\bar g$};
		\node [style=textual] (17) at (-0.75, 0.25) {$\color{blue}{_{q_0 \mapsto (q_1,r)}}$};
		\node [style=textual] (18) at (-0.75, -0) {$_{q_0' \mapsto (q_0',r)}$};
		\node [style=textual] (19) at (-2.75, 0.75) {$\color{blue}{_{q_1 \mapsto (q_2,r)}}$};
		\node [style=textual] (20) at (-2.75, 0.5) {$\color{DeepPink}{_{q_0 \mapsto (q_1,r)}}$};
		\node [style=textual] (21) at (-2.75, 0.25) {$_{q_0' \mapsto (q_0',r)}$};
		\node [style=textual] (22) at (-4, 2) {$\color{blue}{_{q_2 \mapsto (q_3,r)}}$};
		\node [style=textual] (23) at (-4, 1.75) {$\color{DeepPink}{_{q_1 \mapsto (q_2,r)}}$};
		\node [style=textual] (24) at (-4, 1.25) {$_{q_0' \mapsto (q_0',r)}$};
		\node [style=textual] (25) at (-4.5, 3.25) {$\color{DeepPink}_{q_2 \mapsto (q_3,\bar r)}$};
		\node [style=textual] (26) at (-4.5, 3.5) {$\color{blue}_{q_3 \mapsto (q_3,\bar r)}$};
		\node [style=textual] (27) at (-4.5, 2.5) {$_{q_0' \mapsto (q_0',r)}$};
		\node [style=textual] (28) at (-4, 1.5) {$\color{Green}{_{q_0 \mapsto (q_1,r)}}$};
		\node [style=textual] (29) at (-4.5, 3) {$\color{Green}_{q_1 \mapsto (q_2,r)}$};
		\node [style=textual] (30) at (-4.5, 2.75) {$\color{DarkGoldenrod}{_{q_0 \mapsto (q_1,r)}}$};
		\node [style=textual] (31) at (-0.75, -0.25) {$_{p_{\sf EX}, p_{\sf EG}}$};
		\node [style=textual] (32) at (-2.75, -0) {$_{p_{\sf EX}}$};
		\node [style=textual] (33) at (-4, 1) {$_{p_{\sf EX}}$};
		\node [style=textual] (34) at (-4.5, 2.25) {$_{p_{\sf EX}}$};
	\end{pgfonlayer}
	\begin{pgfonlayer}{edgelayer}
		\draw [style=arrow] (1) to node{} (0);
		\draw [style=arrow] (4) to (5);
		\draw [style=arrow] (4) to (6);
		\draw [style=arrow] (3) to (7);
		\draw [style=dashed arrow] (3) to (8);
		\draw [style=blue, transform canvas={yshift=0.3mm,xshift=0.8mm}] (3) to (8);
		\draw [style=pink, transform canvas={yshift=0.8mm,xshift=1.4mm}] (3) to (8);
		\draw [style=green, transform canvas={yshift=-0.3mm,xshift=-0.5mm}] (3) to (8);
		\draw [style=dashed arrow, in=-45, out=135, looseness=0.75] (2) to node{} (3);
		\draw [style=blue, transform canvas={yshift=0.7mm,xshift=0.7mm}] (2) to (3);
		\draw [style=pink, transform canvas={yshift=1.3mm,xshift=1.3mm}] (2) to (3);
		\draw [style=arrow] (2) to (4);
		\draw [style=dashed arrow] (1) to node[right]{} (2);
		\draw [style=blue, transform canvas={yshift=0.7mm,xshift=0.1mm}] (1) to (2);
		\draw [style=dashed arrow] (8) to (9);
		\draw [style=green, transform canvas={xshift=-0.5mm,yshift=-0.1mm}] (8) to (9);
		\draw [style=yellow, transform canvas={xshift=-1mm,yshift=-0.2mm}] (8) to (9);
		\draw [style=arrow] (8) to (10);
		\draw [style=blue, transform canvas={xshift=0.5mm,yshift=-0.1mm}] (8) to (10);
		\draw [style=pink, transform canvas={xshift=1mm,yshift=-0.2mm}] (8) to (10);
		\draw [style=arrow] (7) to (11);
		\draw [style=arrow] (7) to (12);
		\draw [style=arrow] (6) to (13);
		\draw [style=arrow] (6) to (14);
		\draw [style=arrow] (5) to (15);
		\draw [style=arrow] (5) to (16);
	\end{pgfonlayer}
\end{tikzpicture}

%% file: figures/relabeled-rich-tree.tikz
\begin{tikzpicture}
	\begin{pgfonlayer}{nodelayer}
		\node [style=invisible] (0) at (1, 1) {...};
		\node [initial below, style=wn] (1) at (0, 0.5) {$\bar g$};
		\node [style=wn] (2) at (-2, 1) {$g$};
		\node [style=wn] (3) at (-3, 2) {$g$};
		\node [style=wn] (4) at (-1, 2) {$\bar g$};
		\node [style=wn] (5) at (-0.5, 3) {$\bar g$};
		\node [style=wn] (6) at (-1.5, 3) {$g$};
		\node [style=wn] (7) at (-2.5, 3) {$\bar g$};
		\node [style=wn] (8) at (-3.5, 3) {$g$};
		\node [style=wn] (9) at (-3.75, 4) {$g$};
		\node [style=wn] (10) at (-3.25, 4) {$\bar g$};
		\node [style=wn] (11) at (-2.75, 4) {$g$};
		\node [style=wn] (12) at (-2.25, 4) {$\bar g$};
		\node [style=wn] (13) at (-1.75, 4) {$g$};
		\node [style=wn] (14) at (-1.25, 4) {$\bar g$};
		\node [style=wn] (15) at (-0.75, 4) {$g$};
		\node [style=wn] (16) at (-0.25, 4) {$\bar g$};
		\node [style=textual] (17) at (-1, -0) {$\color{blue}{_{q_0 \mapsto (1, q_1), 1 \mapsto r}}$};
		\node [style=textual] (18) at (-3, 0.75) {$\color{blue}{_{q_1 \mapsto (1,q_2), 1 \mapsto r}}$};
		\node [style=textual] (19) at (-3, 0.5) {$\color{DeepPink}{_{q_0 \mapsto (1,q_1)~~~~~}}$};
		\node [style=textual] (20) at (-4.25, 2) {$\color{blue}{_{q_2 \mapsto (1,q_3), 1 \mapsto r}}$};
		\node [style=textual] (21) at (-4.25, 1.75) {$\color{DeepPink}{_{q_1 \mapsto (1,q_2)~~~~~}}$};
		\node [style=textual] (22) at (-4.75, 3.25) {$\color{DeepPink}_{q_2 \mapsto (1,q_3)~~~~~}$};
		\node [style=textual] (23) at (-4.75, 3.5) {$\color{blue}_{q_3 \mapsto (1,q_3), 1 \mapsto \bar r}$};
		\node [style=textual] (24) at (-4.25, 1.5) {$\color{Green}{_{q_0 \mapsto (2,q_1), 2 \mapsto r}}$};
		\node [style=textual] (25) at (-4.75, 3) {$\color{Green}_{q_1 \mapsto (2,q_2), 2 \mapsto r}$};
		\node [style=textual] (26) at (-4.75, 2.75) {$\color{DarkGoldenrod}{_{q_0 \mapsto (3,q_1), 3 \mapsto r}}$};
		\node [style=textual] (27) at (-1, -0.25) {$\color{black}{_{q_0' \mapsto (4, q_0'), 4 \mapsto r}}$};
		\node [style=textual] (28) at (-3, 0.25) {$\color{black}{_{q_0' \mapsto (4, q_0'), 4 \mapsto r}}$};
		\node [style=textual] (29) at (-4.25, 1.25) {$\color{black}{_{q_0' \mapsto (4, q_0'), 4 \mapsto r}}$};
		\node [style=textual] (30) at (-4.75, 2.5) {$\color{black}{_{q_0' \mapsto (4, q_0'), 4 \mapsto r}}$};
		\node [style=textual] (31) at (-1.5, -0.5) {$_{p_{\sf EX}, p_{\sf EG}}$};
		\node [style=textual] (32) at (-3.75, -0) {$_{p_{\sf EX}}$};
		\node [style=textual] (33) at (-5, 1) {$_{p_{\sf EX}}$};
		\node [style=textual] (34) at (-5.5, 2.25) {$_{p_{\sf EX}}$};
	\end{pgfonlayer}
	\begin{pgfonlayer}{edgelayer}
		\draw [style=arrow] (1) to node{} (0);
		\draw [style=arrow] (4) to (5);
		\draw [style=arrow] (4) to (6);
		\draw [style=arrow] (3) to (7);
		\draw [style=dashed arrow] (3) to (8);
		\draw [style=blue, transform canvas={yshift=0.3mm,xshift=0.8mm}] (3) to (8);
		\draw [style=pink, transform canvas={yshift=0.8mm,xshift=1.4mm}] (3) to (8);
		\draw [style=green, transform canvas={yshift=-0.3mm,xshift=-0.5mm}] (3) to (8);
		\draw [style=dashed arrow, in=-45, out=135, looseness=0.75] (2) to node{} (3);
		\draw [style=blue, transform canvas={yshift=0.7mm,xshift=0.7mm}] (2) to (3);
		\draw [style=pink, transform canvas={yshift=1.3mm,xshift=1.3mm}] (2) to (3);
		\draw [style=arrow] (2) to (4);
		\draw [style=dashed arrow] (1) to node[right]{} (2);
		\draw [style=blue, transform canvas={yshift=0.7mm,xshift=0.1mm}] (1) to (2);
		\draw [style=dashed arrow] (8) to (9);
		\draw [style=green, transform canvas={xshift=-0.5mm,yshift=-0.1mm}] (8) to (9);
		\draw [style=yellow, transform canvas={xshift=-1mm,yshift=-0.2mm}] (8) to (9);
		\draw [style=arrow] (8) to (10);
		\draw [style=blue, transform canvas={xshift=0.5mm,yshift=-0.1mm}] (8) to (10);
		\draw [style=pink, transform canvas={xshift=1mm,yshift=-0.2mm}] (8) to (10);
		\draw [style=arrow] (7) to (11);
		\draw [style=arrow] (7) to (12);
		\draw [style=arrow] (6) to (13);
		\draw [style=arrow] (6) to (14);
		\draw [style=arrow] (5) to (15);
		\draw [style=arrow] (5) to (16);
	\end{pgfonlayer}
\end{tikzpicture}

%% file: figures/lean-numbered-tree.tikz
\begin{tikzpicture}
	\begin{pgfonlayer}{nodelayer}
		\node [style=invisible] (0) at (1, 1) {...};
		\node [style=wn, initial below] (1) at (0, 0.5) {$\bar g$};
		\node [style=wn] (2) at (-2, 1) {$g$};
		\node [style=wn] (3) at (-3, 2) {$g$};
		\node [style=wn] (4) at (-1, 2) {$\bar g$};
		\node [style=wn] (5) at (-0.5, 3) {$\bar g$};
		\node [style=wn] (6) at (-1.5, 3) {$g$};
		\node [style=wn] (7) at (-2.5, 3) {$\bar g$};
		\node [style=wn] (8) at (-3.5, 3) {$g$};
		\node [style=wn] (9) at (-3.75, 4) {$g$};
		\node [style=wn] (10) at (-3.25, 4) {$\bar g$};
		\node [style=wn] (11) at (-2.75, 4) {$g$};
		\node [style=wn] (12) at (-2.25, 4) {$\bar g$};
		\node [style=wn] (13) at (-1.75, 4) {$g$};
		\node [style=wn] (14) at (-1.25, 4) {$\bar g$};
		\node [style=wn] (15) at (-0.75, 4) {$g$};
		\node [style=wn] (16) at (-0.25, 4) {$\bar g$};
		\node [style=textual] (17) at (-0.75, 0.25) {$\color{black}{_{1 \mapsto r}}$};
		\node [style=textual] (18) at (-0.75, -0) {$\color{black}{_{4 \mapsto r}}$};
		\node [style=textual] (19) at (-0.75, -0.25) {$_{{\color{blue}{v_{\sf EX}=1}}, v_{\sf EG}=4}$};
		\node [style=textual] (20) at (-2.75, 0.5) {$\color{black}{_{4 \mapsto r}}$};
		\node [style=textual] (21) at (-2.75, 0.75) {$\color{black}{_{1 \mapsto r}}$};
		\node [style=textual] (22) at (-2.75, 0.25) {$_{{\color{DeepPink}{v_{\sf EX}=1}}}$};
		\node [style=textual] (23) at (-3.75, 1.5) {$\color{black}{_{4 \mapsto r}}$};
		\node [style=textual] (24) at (-3.75, 1.25) {$_{{\color{Green}{v_{\sf EX}=2}}}$};
		\node [style=textual] (25) at (-3.75, 2) {$\color{black}{_{1 \mapsto r}}$};
		\node [style=textual] (26) at (-3.75, 1.75) {$\color{black}{_{2 \mapsto r}}$};
		\node [style=textual] (27) at (-4.25, 2.75) {$\color{black}{_{4 \mapsto r}}$};
		\node [style=textual] (28) at (-4.25, 2.5) {$_{{\color{DarkGoldenrod}{v_{\sf EX}=3}}}$};
		\node [style=textual] (29) at (-4.25, 3.5) {$\color{black}{_{1 \mapsto \bar r}}$};
		\node [style=textual] (30) at (-4.25, 3.25) {$\color{black}{_{2 \mapsto r}}$};
		\node [style=textual] (31) at (-4.25, 3) {$\color{black}{_{3 \mapsto r}}$};
	\end{pgfonlayer}
	\begin{pgfonlayer}{edgelayer}
		\draw [style=arrow] (1) to node{} (0);
		\draw [style=arrow] (4) to (5);
		\draw [style=arrow] (4) to (6);
		\draw [style=arrow] (3) to (7);
		\draw [style=dashed arrow] (3) to (8);
		\draw [transform canvas={yshift=0.3mm,xshift=0.8mm}, style=blue] (3) to (8);
		\draw [transform canvas={yshift=0.8mm,xshift=1.4mm}, style=pink] (3) to (8);
		\draw [transform canvas={yshift=-0.3mm,xshift=-0.5mm}, style=green] (3) to (8);
		\draw [style=dashed arrow, in=-45, out=135, looseness=0.75] (2) to node{} (3);
		\draw [transform canvas={yshift=0.7mm,xshift=0.7mm}, style=blue] (2) to (3);
		\draw [transform canvas={yshift=1.3mm,xshift=1.3mm}, style=pink] (2) to (3);
		\draw [style=arrow] (2) to (4);
		\draw [style=dashed arrow] (1) to node[right]{} (2);
		\draw [transform canvas={yshift=0.7mm,xshift=0.1mm}, style=blue] (1) to (2);
		\draw [style=dashed arrow] (8) to (9);
		\draw [transform canvas={xshift=-0.5mm,yshift=-0.1mm}, style=green] (8) to (9);
		\draw [transform canvas={xshift=-1mm,yshift=-0.2mm}, style=yellow] (8) to (9);
		\draw [style=arrow] (8) to (10);
		\draw [transform canvas={xshift=0.5mm,yshift=-0.1mm}, style=blue] (8) to (10);
		\draw [transform canvas={xshift=1mm,yshift=-0.2mm}, style=pink] (8) to (10);
		\draw [style=arrow] (7) to (11);
		\draw [style=arrow] (7) to (12);
		\draw [style=arrow] (6) to (13);
		\draw [style=arrow] (6) to (14);
		\draw [style=arrow] (5) to (15);
		\draw [style=arrow] (5) to (16);
	\end{pgfonlayer}
\end{tikzpicture}